\documentclass[amsmath,amssymb,aps,prl,superscriptaddress,twocolumn]{revtex4-1}

\usepackage{amsfonts,amsmath,amssymb,amsthm}
\usepackage{url}
\usepackage{hyperref}
\hypersetup{colorlinks=false,pdfborder={0 0 0}}

\usepackage{graphicx,color,array}
\usepackage[normalem]{ulem}
\usepackage{times}
\usepackage{subcaption}

\usepackage{mathrsfs}
\usepackage{natbib}
\usepackage{verbatim}
\DeclareMathAlphabet{\mathrsfs}{U}{rsfs}{m}{n}
\DeclareMathAlphabet{\mathpzc}{OT1}{pzc}{m}{it}
\DeclareMathAlphabet{\matheus}{U}{eus}{m}{n}
\DeclareMathAlphabet{\mathbbold}{U}{bbold}{m}{n}

\newcommand{\ba}{\begin{eqnarray}}
\newcommand{\be}{\begin{equation}}
\newcommand{\ee}{\end{equation}}
\newcommand{\beq}{\begin{equation}}
\newcommand{\eeq}{  \end{equation}}
\newcommand{\bea}{\begin{eqnarray}}
\newcommand{\eea}{  \end{eqnarray}}

\newcommand{\ea}{\end{eqnarray}}
\newcommand{\ban}{\begin{eqnarray*}}
\newcommand{\ean}{\end{eqnarray*}}
\newcommand{\Tr}{\operatorname{tr}}
\newcommand{\tr}{\operatorname{tr}}

\newcommand{\ket}[1]{\left|#1\right\rangle}
\newcommand{\bra}[1]{\langle#1|}

\newcommand{\ketbra}[2]{|#1\rangle\langle#2|}
\newcommand{\expect}[1]{\langle#1\rangle}

\newcommand{\eg}{{\it{e.g.}~}}
\newcommand{\ie}{{\it{i.e.}~}}

\newcommand{\rA}{\mathrm{A}}

\newcommand{\bA}{\mathbf{a}}
\newcommand{\rB}{\mathrm{B}}
\newcommand{\rC}{\mathrm{C}}

\newtheorem{theorem}{Theorem}

\begin{document}

\title{General method for constructing local-hidden-variable models for entangled quantum states}

\author{D. Cavalcanti}
\affiliation{ICFO-Institut de Ciencies Fotoniques, The Barcelona Institute of Science and Technology, 08860 Castelldefels (Barcelona), Spain}
\email{daniel.cavalcanti@icfo.es}

\author{L. Guerini}
\affiliation{ICFO-Institut de Ciencies Fotoniques, The Barcelona Institute of Science and Technology, 08860 Castelldefels (Barcelona), Spain}
 \affiliation{Departamento de Matem\'atica, Universidade Federal de Minas Gerais, Caixa
Postal 702, 31270-901, Belo Horizonte, MG, Brazil}

\author{R. Rabelo}
\affiliation{Departamento de Matem\'atica, Universidade Federal de Minas Gerais, Caixa
Postal 702, 31270-901, Belo Horizonte, MG, Brazil}

\author{P. Skrzypczyk}
\affiliation{H. H. Wills Physics Laboratory, University of Bristol, Tyndall Avenue, Bristol, BS8 1TL, United Kingdom}
\affiliation{ICFO-Institut de Ciencies Fotoniques, The Barcelona Institute of Science and Technology, 08860 Castelldefels (Barcelona), Spain}

\begin{abstract}
Entanglement allows for the nonlocality of quantum theory, which is the resource behind device-independent quantum information protocols. However, not all entangled quantum states display nonlocality, and a central question is to determine the precise relation between entanglement and nonlocality. Here we present the first general test to decide whether a quantum state is local, and that can be implemented by semidefinite programming. This method can be applied to any given state and for the construction of new examples of states with local hidden-variable models for both projective and general measurements. As applications we provide a lower bound estimate of the fraction of two-qubit local entangled states and present new explicit examples of such states, including those which arise from physical noise models, Bell-diagonal states, and noisy GHZ and W states.
\end{abstract}


\maketitle
\textit{Introduction.---} Entanglement is one of the defining properties of quantum theory, playing a central role in quantum information science. One of the most astonishing consequences of entanglement is that local measurements on composite quantum systems can produce correlations which are impossible to reproduce by any classical mechanism satisfying natural notions of local causality \cite{BruCavPir14}. Such correlations are the key aspect behind the famous nonlocality of quantum theory, and they are witnessed by the violation of Bell inequalities \cite{Bel64}. Witnessing nonlocality certifies the entanglement of the underlying quantum state in a way which makes no assumptions about the functioning {of} the apparatuses used, a realisation which {led} to the development of the field of device-independent quantum information.

Remarkably, as first shown by Werner \cite{Werner_1989}, although every entangled state needs a quantum channel to be distributed, there exist entangled quantum states whose correlations can be reproduced classically, since they are incapable of displaying nonlocality. More precisely, Werner presented a highly symmetric family of quantum states whose statistics for all possible projective measurements could be reproduced by an ingenious classical model, referred to as a local-hidden-variable (LHV) model.  On the one hand this shows that the relation between entanglement and nonlocality is not straightforward. On the other hand, it shows that not all entangled states are useful for applications in device-independent information processing. Since Werner's original result there have been a number of subsequent results further elucidating the relation between entanglement and nonlocality in terms of finding LHV models for other families of states \cite{Barrett_2002,Ac_n_2006,Almeida_2007,Wiseman_2007,Jevtic_2015,Hirsch_2013} (for a review see \cite{Augusiak_2014}).

Nevertheless, it remains a difficult task to decide whether a given entangled quantum state is nonlocal or not. This lies in the fact that showing that a given state cannot lead to nonlocal correlations requires showing that the statistics of all measurements can be reproduced by a suitable LHV model. Crucially all constructions to date make use of the symmetries present in the quantum states under scrutiny, and consequently they can not be readily applied to other quantum states. In fact,
apart from very recent sufficient condition for the special case of two-qubits (and one-sided projective measurements) \cite{Bowles2015}, there is no general criterion to test whether a given quantum state is local.

Our main contribution here is to present sufficient conditions for a general quantum state to admit a LHV model, either for projective von Neumann measurements, or for general positive-operator-valued measure (POVM) measurements, that can be tested via semi-definite programming (SDP), an efficient form of convex optimisation that can be readily implemented in practice. We also show how this method can be modified to provide a means to randomly generate local quantum states.
We show the power of these tests by providing a lower-bound estimate on the volume of the set of entangled two-qubit states that possess LHV models for projective and POVM measurements, and by presenting several examples of new local entangled states, including those that would arise from local amplitude-damping noise, two-qubit Bell diagonal states, and three-qubit noisy GHZ and W states. 
Our method focuses on a particular class of LHV models, known as local-hidden-state (LHS) models, which naturally arise in the context of quantum steering \cite{Schr_dinger_1936,Wiseman_2007}, a closely related concept to nonlocality.An advantage of such models is that they automatically imply a LHV model when one of the parties apply POVM measurements. A disadvantage is that there exist entangled states that admit LHV models but do not admit LHS models \cite{Wiseman_2007,Quintino_2015}. As we will see in what follows, even with this restriction, our tests are still strong enough to find models for many interesting states.

\textit{Preliminaries.---}Let us start by defining more precisely LHV and LHS models. Suppose that Alice and Bob apply local measurements defined by measurement operators $\{M_{a|x}\}$ and $\{M_{b|y}\}$ ($x$ and $y$ labels measurement choices and $a$ and $b$ outcomes) on a shared state $\rho_{\rA\rB}$. The set of conditional probability distributions they observe is
\begin{equation}
\label{eqn:behaviour}
P(a,b|x,y) = \tr\left[\left(M_{a|x}\otimes M_{b|y}\right)\rho_{\rA\rB}\right].
\end{equation}
The state $\rho_{\rA\rB}$ is said to have a LHV model for these measurements if $P(a,b|x,y)$ can be written as
\begin{equation}
\label{eqn:lhv-model}
P(a,b|x,y)  = \int d\lambda q(\lambda) P(a|x,\lambda)P(b|y,\lambda)
\end{equation}
where $\lambda$ is the so-called \emph{shared {local} hidden variable} and $\int d\lambda q(\lambda) = 1$. This decomposition can be thought as coming from the following model: a classical variable $\lambda$ is randomly chosen according to the probability density $q(\lambda)$ and sent to Alice and Bob. Upon receiving $\lambda$ and choosing their measurement, Alice and Bob output $a$ and $b$ according to the distributions $P(a|x,\lambda)$ and $P(b|y,\lambda)$, respectively. Of particular interest is when the sets $\{M_{a|x}\}$ and $\{M_{b|y}\}$ contain either all projective measurements, or all POVM measurements.

A subclass of LHV models is that of local-hidden-state (LHS) models. Let us consider now that only Alice measures $\rho_{\rA\rB}$. The (unnormalised) state on Bob's side, conditioned on Alice having observed the outcome $a$ of measurement $x$ is
\begin{equation}
\label{eqn:assemblage}
\sigma_{a|x}=\tr_\rA\left[\left(M_{a|x}\otimes \mathbb{I}_\rB \right)\rho_{\rA\rB}\right],
\end{equation}
where $\tr[\sigma_{a|x}]=P(a|x)$ is the probability that Alice obtains the outcome $a$. If these post-measurement states can be written in the form
\begin{equation}
\label{eqn:lhs-model}
\sigma_{a|x}=\int d\lambda q(\lambda)P(a|x,\lambda)\rho_\lambda,
\end{equation}
where $\rho_\lambda\geq0$, $\tr\rho_\lambda=1$ for all $\lambda$, and $\int d\lambda q(\lambda) = 1$, $\rho_{\rA\rB}$ is said to have a LHS model for these measurements. It can be easily checked that if Bob measures his share of the state  \eqref{eqn:lhs-model} with any set of POVM measurements, the probability distributions observed will have the form \eqref{eqn:lhv-model}. This means that the existence of a LHS model implies a LHV model for arbitrary POVM measurements on Bob's side. Note that LHS models are not as powerful as general LHV models; there exist states that provably have an LHV model but no LHS model \cite{Wiseman_2007,Quintino_2015}.

\textit{Main Results.---} The main insight behind the following theorems is to replace the problem of finding a LHS model for a physical state and an \emph{infinite set of measurements}, to the one of finding a model for a \emph{non-physical} operator and a \emph{finite set of measurements}. As we will discuss afterwards, this is a huge simplification that will allow us to test for LHS models via SDP. 

\begin{theorem} [LHS model for projective measurements]
Let $\mathcal{M}$ be a finite collection of projective measurements in $\mathbb{C}^{d_\rA}$. A state $\rho_{\rA\rB}$ acting on $\mathbb{C}^{d_\rA}\otimes \mathbb{C}^{d_\rB}$ admits a LHS model for \emph{all} projective measurements if there exists a unit-trace operator $O_{\rA\rB}$ acting on the same Hilbert space, such that $O_{\rA\rB}$ admits a LHS model for the measurements in $\mathcal{M}$, and
\begin{equation}\label{mala}
\rho_{\rA\rB} = rO_{\rA\rB} + (1-r)\frac{\mathbb{I}_\rA}{d_\rA}\otimes O_{\rB},
\end{equation}
where $r$ is the radius of the insphere\footnote{The insphere of a polytope is the largest centered sphere contained in it.} of the polytope generated by $\mathcal{M}$.
\end{theorem}

Here we prove this theorem for the case of $d_\rA=2$. A proof for arbitrary $d_\rA$ can be found in the Appendix.

\begin{figure}
\centering
\includegraphics[width=0.6\columnwidth]{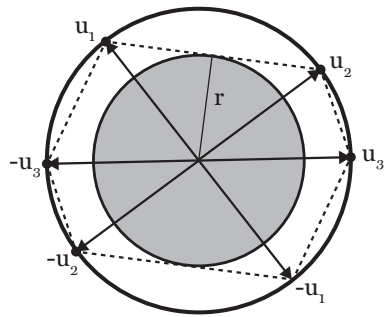}
\caption{\label{Fig1} Diagrammatic representation of the method (restricted to two-dimensions for illustrative purposes). The vectors $\pm\hat{u}_x$ are the Bloch vectors corresponding to each of the measurements from the set $\mathcal{M}$. The area enclosed by the dashed lines is the polytope that these measurements form. Any measurement contained inside this polytope can be simulated by appropriately mixing the LHS model that simulates the measurements in $\mathcal{M}$. The shaded circle, of radius $r$ is the largest circle which is completely contained in the convex hull, and contains all noisy projective measurements $\Pi_{a|\hat{u}}^{(r)}$.%
}
\end{figure}

\begin{proof}
Let $\mathcal{M}$ define a finite set of measurements for Alice given by measurement operators $\Pi_{a|\hat{u}_x} = \frac{\mathbb{I}+(-1)^a \hat{u}_x \dot~\vec{\sigma}}{2}$, where $x=1,..,m_{\rA},\ a,b$ $\in \{0,1\}$, and $\vec{\sigma} = (\sigma_\mathrm{x}, \sigma_\mathrm{y}, \sigma_\mathrm{z})$ is the vector of Pauli operators and $\hat{u}$ a three dimensional unit vector. This measurement set can be chosen arbitrarily -- for example in a regular fashion (along the vertices or faces of a regular solid), or at random. Suppose that these measurements, when applied to a given operator $O_{\rA\rB}$, have a LHS description of the form \eqref{eqn:lhs-model},
\begin{equation}
\tr_\rA\left[\left(\Pi_{a|\hat{u}_x}\otimes \mathbb{I}_\rB \right)O_{\rA\rB}\right]=\int d\lambda q(\lambda)P(a|\hat{u}_x, \lambda)\rho_\lambda, ~\forall~a,x.
\end{equation}

Note that any set of measurements that can be performed as a convex combination of the measurements in $\mathcal{M}$ also has an LHS description. This is valid, in particular, for noisy von Neumann measurements whose elements are contained within a shrunken Bloch sphere completely contained inside the convex hull of $\mathcal{M}$  (see Fig.~\ref{Fig1}). This sphere is given by depolarized measurement operators $\Pi_{a|\hat{u}}^{(r)} = r \Pi_{a|\hat{u}} + (1-r) \mathbb{I}_\rA/2$, where $r$ is the radius of the insphere of the polytope generated by the convex hull of $\mathcal{M}$.

Finally notice that
\begin{equation}
\tr_\rA[(\Pi_{a|\hat{u}}^{(r)}\otimes \mathbb{I}_\rB) O_{\rA\rB} ] = \tr_\rA[( \Pi_{a|\hat{u}}\otimes \mathbb{I}_\rB) \rho_{\rA\rB} ],
\end{equation}
assuming that $\rho_{\rA\rB} = rO_{\rA\rB}+(1-r)\frac{\mathbb{I}_\rA}{2}\otimes O_{\rB}$.
That is, applying noisy measurements on an operator $O_{\rA\rB}$ is equivalent, at the level of the states prepared for Bob, to applying noise-free measurements on a noisy version of $O_{\rA\rB}$, denoted here as $\rho_{\rA\rB}$. Therefore, if $O_{\rA\rB}$ admits a LHS model for the set $\mathcal{M}$, then it also does for the  set $\{\Pi_{a|\hat{u}}^{(r)}\}$, which implies that $\rho_{\rA\rB}$ admits a LHS model for all projective measurement{s}.
\end{proof}

Note first that the operator $O_{\rA\rB}$ need not to be a valid density operator (it can have negative eigenvalues). The requirements on $O_{\rA\rB}$  are that it has unit trace, admits a LHS model for the measurements in $\mathcal{M}$, and that it becomes equal to $\rho_{\rA\rB}$ when depolarized. Note also that in the case that $\mathcal{M}$ is the (infinite) set of all projective measurements, then this is precisely a brute force test for the existence of a LHS model. Thus, our method can be seen to provide {a} sequence of tests (sufficient conditions), in terms of the set $\mathcal{M}$, for a state to have an LHS model, which in the limit converges to the brute force test.

To further generalise this result to accommodate general POVMs we can make use of a result from Ref.~\cite{Hirsch_2013}, that if $\rho_{\rA\rB}$ has a LHS model for projective measurements, than the state $\rho_{\rA\rB}'=(1/d_\rA)\rho_{\rA\rB}+(1-1/d_\rA)\gamma_\rA\otimes\rho_\rB$ has a LHS model for all POVMs, where $d_\rA$ is the local Hilbert space dimension of Alice and $\gamma_\rA$ is an arbitrary state\footnote{In fact, \cite{Hirsch_2013} present a more general construction which works for LHV models. We only need the weaker result stated, which is implicit in their construction.}. Combining this result with the above theorem, we obtain the following

\begin{theorem}  [LHS for POVM measurements]
A state $\rho_{\rA\rB}$ acting in $\mathbb{C}^{d_\rA}\otimes \mathbb{C}^{d_\rB}$ admits a LHS model for \emph{all} POVMs if there exists an operator $O_{\rA\rB}$ that admits a LHS model for $\mathcal{M}$ such that
\begin{equation}\label{devendra}
\rho_{\rA\rB} = \frac{1}{d_\rA}\left[ r O_{\rA\rB}+(1-r)\frac{\mathbb{I}_\rA}{d_\rA}\otimes O_{\rB} \right]+\frac{d_\rA-1}{d_\rA}\gamma_\rA \otimes O_{\rB},
\end{equation}
where $\gamma_\rA$ is an arbitrary state.
\end{theorem}

Note however, that unlike in the previous case, which became a brute force search for the existence of an LHS model for all projective measurements in an appropriate limit, this test provides only a sufficient criteria.

Both theorems can be easily adapted to the case of LHV models by applying the same ideas also to Bob's system \cite{Geneva}. That is, one can also choose a set of measurements to Bob, compute the corresponding radius $r_B$, impose that Alice{'s} and Bob's measurements generate local probability distributions and locally depolarise according to Alice and Bob's shrinkng factors.

\textit{SDP formulation.---}We now provide explicit SDP formulations of Theorems 1 and 2. We start by choosing a finite set of measurements $\mathcal{M}$ and calculating $r$, given by the distance between the closest facet of the polytope generated by $\mathcal{M}$ and the origin, which can easily be computed by standard vertex enumeration algorithms \cite{porta, lrs}.
Since $\mathcal{M}$ is finite, we can restrict to a finite set of hidden variables \cite{Pusey_2013} when imposing an LHS model for the operator $O_{\rA\rB}$. Without loss of generality we take $\lambda = \lambda_1\cdots\lambda_{m_\rA}$ to be a $m_\rA$-length bit-string, which specifies a (deterministic) outcome for each of the $m_\rA$ measurements of Alice: $a = \lambda_x$ when the measurement along direction $\hat{u}_x$ is performed. There are $d^{m_\rA}$ distinct deterministic specifications. Thus, according to Theorem 1, the following SDP tests for a LHS model for projective measurements on the state $\rho_{\rA\rB}$:
\begin{align}
\mathrm{given}&& &\rho_{\rA\rB}, \mathcal{M}, r \nonumber \\
\mathrm{find}&&  &O_{\rA\rB},\{\rho_\lambda\}_\lambda  \nonumber \\
\mathrm{s.t.}&& &\tr_\rA\left[\left(\Pi_{a|\hat{u}_x}\otimes \mathbb{I}_\rB \right)O_{\rA\rB}\right]=\sum_\lambda D_\lambda(a|x)\rho_\lambda, \quad\forall~a,x  \nonumber  \\
 && &\rho_\lambda\geq0, \quad\forall~\lambda \label{eqn:sdp-proj-fixed-state} \\
 && & rO_{\rA\rB}+(1-r)\frac{\mathbb{I}_\rA}{d}\otimes O_{\rB}=\rho_{\rA\rB}, \nonumber
\end{align}
where 
$D_\lambda (a|x) = \delta_{a,\lambda_x}$ are deterministic response functions. 

Following Theorem 2 we can substitute the last constraint in the above SDP by Eq.~(\ref{devendra}) to test, with a given $\gamma_\rA$, for the existence of an LHS model for all POVM measurements.
These programs can also be adapted to test families of states $\rho(w)$ that depend linearly on a parameter $w$ (\eg Werner states): instead of running the feasibility problem \eqref{eqn:sdp-proj-fixed-state} one can maximise (or minimise) $w$ subject to the same constraints. This finds the value $w^*$ such that for all $w \leq w^*$, states within the family have an LHS model.

\textit{Extensions.---}The previous methods extend to multipartite states in a rather straightforward way. In particular, extending $\rB \to \rB_1\otimes \cdots \otimes \rB_k$, we demand in addition that each $\rho_\lambda$ in \eqref{eqn:sdp-proj-fixed-state} (now an operator on $\mathcal{H}_{\rB_1} \otimes \cdots \otimes \mathcal{H}_{\rB_k}$) is a fully separable state. This is easily seen to provide an LHV model where each Bob can perform arbitrary POVM measurements. Note that although imposing separability is in general difficult; for the case where Bob holds two qubits, imposing positive partial transpose (PPT) is sufficient. 
In the case of higher dimensional systems, although in principle the above method still applies, the number of measurements necessary to generate a polytope with a large insphere grows quickly with $d_\rA$ (which is necessary to keep the amount of noise low). This implies that the above SDPs become too costly to be used in practice.

\textit{Example 1.---} As an illustration of the technique we first investigate the Bell diagonal states, given by
$\rho_\mathrm{Bell}=\sum_i p_i \ket{\Psi_i}\bra{\Psi_i}$,
where $\ket{\Psi_i}$ are the four Bell states, $p_i\geq0$, and $\sum_i p_i=1$ have LHS models. In this case we adapted the SDP \eqref{eqn:sdp-proj-fixed-state} to maximise $p_1$ provided the same constraints. We find $p_1\approx0.4454$, and $p_2=p_3=p_4=(1-p_1)/3$, which is a Werner state, using $\mathcal{M}$ along the vertices of the rhombicuboctahedron, an Archimedian solid with 24 vertices. Notice that the analytical construction of Werner \cite{Werner_1989} provides a model for $p_1 \leq 1/2$, thus with 12 measurements our method already recaptures $\approx 89 \%$ of LHS Werner states. We also looked at rank-3 Bell diagonal states, by setting $p_4=0$, and found the largest $p_1$ equal to $0.5664$, with the same $\mathcal{M}$.

\textit{Example 2.---} As a more physical example we consider an initial maximally entangled state $\ket{\Phi^+} = (\ket{00} + \ket{11})/\sqrt{2}$ undergoing independent local amplitude damping given by the evolution $\rho(t)=\sum_{i,j}E_i\otimes E_j \rho(0)E_i^\dag\otimes E_j^\dag$, defined by the Kraus operators $E_0=\ketbra{0}{0}+\sqrt{1-e^{-\gamma t}}\ketbra{1}{1}$ and $E_1=\sqrt{e^{-\gamma t}}\ketbra{0}{1}$. This noisy model is used to describe spontaneous decay of two-level systems \cite{Nielsen:2011:QCQ:1972505} and is particularly relevant for atomic Bell experiments \cite{Hofmann72,2010Natur.464.1021P}. While the evolved state becomes separable only asymptotically (\ie for $t\rightarrow\infty$), we found it to have an LHS model for all $\gamma t\gtrsim -\ln 0.60$. 

\textit{Example 3.---} Finally, we consider noisy 3-qubit GHZ and W states given $\rho(p) = p \ketbra{\psi}{\psi} + (1-p)\mathbb{I}/8$, where  $\ket{\psi} = \ket{\mathrm{GHZ}} := (\ket{000} + \ket{111})/\sqrt{2}$ or $\ket{\psi} = \ket{W} := (\ket{001} + \ket{010} + \ket{100})/\sqrt{3}$. These states are fully separable for $p \leq 0.2$ and $p \leq 0.2096$ respectively. With $\mathcal{M}$ corresponding to the rhombicuboctahedron we found that these states are LHS for projective measurements for $p \leq 0.232$ and $p \leq 0.228$ respectively.

\textit{Generating entangled states with LHS models.---}
A complementary problem to the one of deciding if a target state is local, is to generate local entangled states.
Furthermore, it is also interesting generating local states which contain as much entanglement as possible. To this end, we make use of the concept of entanglement witnesses.

Entanglement witnesses are Hermitian operators $W$ for which a negative expectation value for the state $\rho_{\rA\rB}$, $\Tr[W\rho_{\rA\rB}]<0$, certifies that it is entangled. As shown in Refs.~\cite{Brand_o_2005,Eisert_2007}, if $W$ has additional appropriate structure, the absolute value of this negative expectation value also provides a lower bound on the amount of entanglement of $\rho_{\rA\rB}$, \ie $E(\rho_{\rA\rB})\geq -\Tr[W\rho_{\rA\rB}]$. Finally, such entanglement witnesses themselves can be obtained through simple SDPs, where by imposing the different constraints on $W$ we obtain  (see Appendix) estimators for different entanglement quantifiers $E(\rho_{\rA\rB})$ \cite{Brand_o_2005,Eisert_2007} 

We now propose a method to generate entangled states with LHS models and high entanglement. We start with a given witness $W$ (obtained via an SDP). As before, we choose a set of measurements $\mathcal{M}$ and compute the radius of the insphere $r$. We now search for the state which maximally violates the witness and has a LHS model for projective measurements by solving the following SDP:
\begin{align}
 \min_{O_{\rA\rB},\rho_\bA}&& & \tr[W(rO_{\rA\rB}+(1-r)\frac{\mathbb{I}_\rA}{d}\otimes O_{\rB})]  \nonumber  \\
 \mathrm{s.t.}&& &\tr_\rA\left[\left(\Pi_{a|\hat{u}_x}\otimes \mathbb{I}_\rB \right)O_{\rA\rB}\right]=\sum_\mathrm{a} D_\mathrm{a}(a|x)\rho_\mathrm{a}, \forall~a,x  \nonumber  \\
 && &\rho_\mathrm{a}\geq0, \quad\forall~\mathrm{a},\quad\quad \tr[O_{\rA\rB}] = 1, \label{eqn:sdp-proj-random-state}  \\
 && & rO_{\rA\rB}+(1-r)\frac{\mathbb{I}_\rA}{d}\otimes O_{\rB}\geq 0. \nonumber  
\end{align}
If the solution of this SDP is negative, then the minimising operator $\rho_{\rA\rB}^*=rO_{\rA\rB}^*+(1-r)\frac{\mathbb{I}_\rA}{d}\otimes O^*_{\rB}$ is an entangled state which has a LHS model: entanglement is guaranteed by the violation of the witness and the fact it is has a LHS model is imposed by the constraints of the SDP.

Once we find an example of a LHS entangled state $\rho_{\rA\rB}^*$, we can iterate this procedure and find new examples with more entanglement: we find the entanglement witness $W^*$ which is optimal for the state $\rho_{\rA\rB}^*$ and use $W^*$ in the SDP \eqref{eqn:sdp-proj-random-state} to find a new state $\rho^{**}_{\rA\rB}$, which is generally more entangled according to the chosen quantifier. This procedure can then be iterated until it converges \footnote{Typically this is convergence up to numerical precision. Note also that this can occur at a local maximum.}. Note that different quantifiers of entanglement have different properties, and thus exploring a number of different quantifiers can provide LHS states with different properties. Finally, as before, we can adapt \eqref{eqn:sdp-proj-random-state} accordingly to Eq.~\eqref{devendra} to find examples of entangled states with LHS models for all POVM measurements.

Using this method we generated a large list of bipartite entangled states which have LHS models for projective and POVM measurements \footnote{The data will be maintained at \url{https://git.io/vV7Bu}}. In the Appendix we make an analysis of these examples in terms of their entanglement content and other relevant parameters. Finally, by using entanglement witnesses that detect genuine multipartite entanglement \cite{Guhne} we were also able to obtain new examples of genuine tripartite entangled three-qubit states with LHS models for projective measurements. To the best of our knowledge, only two examples were previously known \cite{GezaToni,MultiGeneva}.

\textit{Estimating the volume of LHS states.---} The previous programs can be directly applied to lower bound the relative volume of the set of entangled states that admit LHS models. We uniformly sampled $2\times10^{4}$ two-qubit states according to the Hilbert-Schmidt and Bures measures, for which we obtained, $\approx23\%$ and $\approx7\%$ separable states, respectively, in good accordance with the values $24.2\%$ and $7.3\%$, obtained from geometrical arguments \cite{Slater}.  We then applied the above SDPs to estimate how many of the entangled states admit LHS models \footnote{All the SDPs implemented in this work used {\sc matlab} and the packages {\sc cvx} \cite{cvx,gb08} and {\sc qetlab} \cite{qetlab}.}. With the measurements $\mathcal{M}$ chosen to be the vertices of the icosahedron ($r \approx 0.79$), we obtain that ${\gtrsim} 25\%$ of the entangled states sampled according to the Hilbert-Schmidt measure are LHS, while ${\gtrsim} 7\%$ are LHS using the Bures measure. We were not able to obtain any entangled state admiting LHS models for POVMs by applying the same technique with measurements given by the icosahedron. A better estimation of the volume of the set of local states could be obtained, both for projective measurements and POVMs, by considering more measurements in the set $\mathcal{M}$.



\textit{Discussion.---} Not all entangled quantum states exhibit nonlocality -- the strongest signature of their inseparability. Understanding the relation between nonlocality and entanglement is an important problem, and it has been notoriously difficult to find general purpose methods for determining which entangled states are local. In this paper we have presented a criterion for a state to admit a LHS model for projective or general measurements. Although LHS models are only a subset of general LHV models, we have demonstrated the power of our criteria by finding new physical examples of multipartite entangled states that are local.

We also showed how our method naturally provides a method to generate examples of entangled local states, and used it to give the first estimate of the relative volume of the set of entangled two-qubit states that admit LHS models, showing that a significant fraction of them in fact do so. As a consequence, our results lower bound the fraction of states useless as resources for any device-independent quantum information processing task.

Our method works particularly well for projective measurements on two- or multi-qubit states, becoming equal to a brute force search in the appropriate limit. In \cite{Geneva} it was further shown how a variant of the above methods allow provide necessary and sufficient criteria for general POVM measurements in the limit. In the future it would be interesting to build upon the general methods presented here to provide practical tests for higher dimensional and multipartite systems.

\textit{Note added.---}During the development of this work we became aware of a complementary work by F. Hirsch et al. \cite{Geneva}.

\acknowledgements
\textit{Acknowledgements.---}We thank M.T. Quintino, F. Hirsch and N. Brunner for discussions. We specially thank F. Hirsch for pointing out the possibility of using non-physical operators in the methods developed. LG is thankful to the Lepant Institute of Quantum Information and Decoherence for the warm hospitality. This work was supported by the Beatriu de Pin\'os fellowship (BP-DGR 2013), the Spanish MINECO (Severo Ochoa Grant SEV-2015-0522 and FOQUS FIS2013-46768-P), the Generalitat de Catalunya (SGR875 and SGR875), the ERC AdG NLST, the ERC CoG QITBOX, CAPES (Brazil) and the Brazilian program Science Without Borders.

\begin{appendix}
\section{Appendix 1. Generalization of the method for dimensions $d>2$}

Here we show that the constructions presented in the main text hold for any finite dimension $d$. We start by noticing that every $d$-dimensional Hermitian operator with unit trace can be written as
\begin{equation}
\frac{\mathbb{I}_d + \vec{v}\cdot \vec{\lambda}}{d},
\end{equation}
where $\mathbb{I}_d$ denotes the identity operator acting on $\mathbb{C}^d$, $\vec{v}$ is a generalized Bloch vector lying in $\mathbb{R}^{d^2-1}$ and $\vec{\lambda}=(\lambda_1,...,\lambda_{d^2-1})$ is the vector of generalized Gell-Mann matrices\footnote{The generalized Gell-Mann matrices generalize also the Pauli matrices.}, that together with the identity form a basis for the space of Hermitian operators \cite{Bertlmann}. In this basis, every rank-1 projector acting on $\mathbb{C}^d$ can be associated to a unit vector, and therefore every rank-1 projective measurement is associated to a set of unit vectors. Notice that we can restrict to rank-1 projective measurements (\ie with $d$ outcomes), making use of the Spectral Theorem and coarse graining.

One difference from the qubit case is that for arbitrary $d$ there are unit vectors that are not associated to projectors \cite{Bertlmann}. However, the collection of projectors being mapped to a subset of the unit sphere in $\mathbb{R}^{d^2-1}$ is enough for our purposes. Indeed, if a shrunken version of the unit sphere fits inside a polytope then a noisy version of all projective measurements does also, and we can decompose them as convex combinations of the extremal points of such a polytope.

Another alteration is that for $d>2$ we interpret each qudit measurement as a vector of measurement operators $M=(M_1, ..., M_d)$. Therefore, for a measurement $N$ to be inside the polytope generated by the subset $\mathcal{M}$ of measurements means to be written as a convex combination
\begin{equation}
  (N_1,...,N_d) = \sum_{M\in\mathcal{M}}{p_M (M_1,...,M_n)},
\end{equation}
where $p_M$ is a probability vector.
This is equivalent to
\begin{equation}
  N_i = \sum_{M\in\mathcal{M}}{p_M M_i},
\end{equation}
for $i=1,...,d$. If $\mathcal{M}$ is composed by projective measurements, that is to say that the vector associated to $N_i$ lies inside the polytope generated by the vector representation of $\{M_i\}_{M\in\mathcal{M}}$, which is inscribed in the unit sphere, for each $i=1,...,d$.

As mentioned in the paper, the SDP (10) and its variations make use of $d^{m_\rA}$ deterministic response functions to find the LHS model for $\rho_{\rA\rB}$, where $m_\rA$ is the number of measurements in $\mathcal{M}$. Since we are dealing with a sphere in $\mathbb{R}^{d^2-1}$, it takes at least $d^2$ measurements in $\mathcal{M}$ to generate a polytope with non-zero volume. This leads to $d^{d^2}$ deterministic strategies in the SDP (10), which already for $d=3$ becomes prohibitively expensive for a desktop computer in terms of computational power.

\section{Entanglement quantifiers and witnesses}
In this appendix we discuss the use of entanglement witnesses as lower bounds on entanglement quantifiers and the fact that finding optimal witnesses can be approximated by solving SDPs. The list the entanglement quantifiers and associated witnesses that we used in our second method for finding random examples of quantum states can be found in Table \ref{tab:witnesses}.

In Ref. \cite{Brand_o_2005} it was shown that many entanglement quantifiers can be written as
\begin{equation}\label{eqn:witnessed-entanglement}
E(\rho)=-\min_{W\in \mathcal{W}_E} \tr(W\rho),
\end{equation}
where $\mathcal{W}_E$ is the set of entanglement witnesses satisfying some additional properties, dependent on the choice of quantifier $E(\cdot)$.  In some cases, the above minimisation problem \eqref{eqn:witnessed-entanglement} is natively an SDP, or can be approximated by one. This is the case, for instance, for the robustness-based quantifiers, which determine how much noise can be added to a quantum state before it becomes separable. What is important to us is that solving this for a given state provides the optimal entanglement witness for that state. We can thus use such a witness as the starting point of our method 2 (SDPs (9) and (10) of the main text).

In the present study we have chosen 7 such quantifiers, summarised in Table \ref{tab:witnesses}
\begin{table*}[h!]
    \begin{tabular}{p{2cm}|c|c|c}
    Quantifier & Given by &Primal & Dual \\ \hline
    One-sided random robustness & {$E_\mathrm{1SRR} = \frac{\mu^*}{1+\mu^*}$}
    & $\begin{aligned} &\mu^* = \min \mu \\
    &\mathrm{s.t } \quad \mu \geq 0, \\
    &(\rho_{\rA\rB}+ \mu \mathbb{I}_\rA/d_\rA\otimes \rho_\rB)\in \mathcal{S}
    \end{aligned}$
    & $\begin{aligned}&\mu^* = -\min \Tr[W\rho_{\rA\rB}] \\
    &\mathrm{s.t.} \quad W\in \mathcal{W}, \\ &\Tr[W(\mathbb{I}_\rA/d_\rA\otimes\rho_\rB)]\leq 1
    \end{aligned}$
    \\ \hline
    One-sided fixed robustness & {$E_\mathrm{1SFR} = \frac{\mu^*}{1+\mu^*}$}
    & $\begin{aligned} &\mu^* = \min \mu
    \\&\mathrm{s.t.} \quad \mu \geq 0, \\
    &(\rho_{\rA\rB}+ \mu \gamma_\rA\otimes \rho_\rB)\in \mathcal{S}
    \end{aligned}$
    & $\begin{aligned}&\mu^* = -\min \Tr[W\rho_{\rA\rB}]\\
    &\mathrm{s.t.}\quad W\in \mathcal{W}, \\
    & \Tr[W(\gamma_\rA\otimes\rho_\rB)]\leq 1
    \end{aligned}$
    \\ \hline
    One-sided generalised robustness & {$E_\mathrm{1SGR} = \frac{\mu^*}{1+\mu^*}$}
    & $\begin{aligned}&\mu^* = \min \Tr[\omega_\rA]\\
    &\mathrm{s.t.}\quad \omega_\rA \geq 0, \\
    &(\rho_{\rA\rB}+ \omega_\rA\otimes \rho_\rB)\in \mathcal{S}
    \end{aligned}$
    & $\begin{aligned}&\mu^* = -\min \Tr[W\rho_{\rA\rB}]\\
    &\mathrm{s.t.}\quad W\in \mathcal{W}, \\
     &\Tr_\rB[W(\mathbb{I}_\rA\otimes\rho_\rB)]\leq \mathbb{I}_\rA
     \end{aligned}$
    \\ \hline
    Random robustness & {$E_\mathrm{RR} = \frac{\mu^*}{1+\mu^*}$}
    & $\begin{aligned}&\mu^* = \min \mu \\
    &\mathrm{s.t.}\quad \mu \geq 0, \\
    &(\rho_{\rA\rB}+ \mu \mathbb{I}_{\rA\rB}/4)\in \mathcal{S}
    \end{aligned}$
    & $\begin{aligned}&\mu^* = -\min \Tr[W\rho_{\rA\rB}]\\
    &\mathrm{s.t.}\quad W\in \mathcal{W}, \\
    &\Tr[W]\leq 4
    \end{aligned}$
    \\ \hline
    Generalised Robustness & {$E_\mathrm{GR} = \frac{\mu^*}{1+\mu^*}$}
    & $\begin{aligned} &\mu^* = \min \tr[\omega_{\rA\rB}]\\
    &\mathrm{s.t.} \quad\omega_{\rA\rB} \geq 0, \\
    &(\rho_{\rA\rB}+ \omega_{\rA\rB})\in \mathcal{S}
    \end{aligned}$
    & $\begin{aligned}&\mu^* = -\min \Tr[W\rho_{\rA\rB}]\\
    &\mathrm{s.t.}\quad W\in \mathcal{W}, \\
    &W\leq \mathbb{I}_{\rA\rB}
    \end{aligned}$
    \\ \hline
    Best separable approximation & {$E_\mathrm{BSA} = \mu^*$}
    & $\begin{aligned}&\mu^* = 1-\max \Tr[\omega_{\rA\rB}]\\
    &\mathrm{s.t.}\quad \omega_{\rA\rB}\in \mathcal{S}, \\
    &\rho_{\rA\rB}\geq \omega_{\rA\rB}
    \end{aligned}$
    & $\begin{aligned}&\mu^* = -\min \Tr[W\rho_{\rA\rB}]\\
    &\mathrm{s.t.}\quad W\in \mathcal{W}, \\
    &-W \leq \mathbb{I}_{\rA\rB}
    \end{aligned}$
    \\ \hline
    Negativity & {$E_\mathrm{Neg} = \mu^*$}
    & $\begin{aligned}&\mu^* = \min \Tr[\omega_{\rA\rB}]\\
    &\mathrm{s.t.} \quad\omega_{\rA\rB} \geq 0, \\
    &\rho_{\rA\rB}^{\Gamma_\rA}+ \omega_{\rA\rB} \geq 0
    \end{aligned}$
    & $\begin{aligned}&\mu^* = -\min \Tr[W\rho_{\rA\rB}]\\
    &\mathrm{s.t.} \quad W^{\Gamma_\rA} \geq 0, \\
    &W^{\Gamma_\rA}\leq \mathbb{I}_{\rA\rB}
    \end{aligned}$
    \label{tab:witnesses}
    \end{tabular}
\caption{\label{tab:witnesses} The 7 quantifiers of entanglement used to generate entanglement witnesses for our randomised search. All quantifiers are such that $E = 0$ for separable states, and $E > 0$ quantifies entanglement. The first five are robustness-type quantifiers, with the specific definition obvious from the primal representation. For the one-sided fixed robustness, $\gamma_{A} $ is chosen to be the projector onto the state $\ket{0}$. The last two are the best separable approximation -- the minimal admixture of entangled state necessary in any decomposition of the state; and the negativity.}
\end{table*}

As an example, for the one-sided random robustness, the set $\mathcal{W}_{E_\mathrm{1SRR}}$, which also depends on the state $\rho_{\rA\rB}$ under consideration, is $\mathcal{W}_{E_\mathrm{1SRR}} = \{W | W \in \mathcal{W}, \tr[W(\mathbb{I}/d_\rA \otimes \rho_\rB)] \leq 1\}$.

Note that in general the condition in the primal problem $\sigma_{\rA\rB} \in \mathcal{S}$, that the state $\sigma_{\rA\rB}$ should be a separable state, and similarly the condition in the dual problem $W\in \mathcal{W}$, that $W$ should be a valid entanglement witness, are complicated constraints. More precisely, there is no efficient way to strictly enforce that a state is separable or that an operator is an entanglement witness, apart from in the simple case of qubit-qubit or qubit-qutrit systems. In these cases separability is equivalent to positivity under partial transposition (PPT), and the above optimisation problems become SDPs, which are then readily solved using standard software packages.

In the general case, instead of calculating directly a given entanglement quantifier, we instead find lower bounds by relaxing the separability constraint. In particular, we can relax this to PPT, or more generally to the set of states that admit a $k$-symmetric PPT extension \cite{Doherty1,Doherty2}. Both constraints can be implemented efficiently via SDP, and are weaker than imposing separability. Hence in both cases we can efficiently obtain lower bounds on the quantifiers. Consequently, the entanglement witness that is obtained from the dual, which we stress is still a valid entanglement witness, it not only positive on separable states, but also positive on all PPT states, or on all states that admit an extension, respectively.

\section{Appendix 2. Analysis of examples of LHS states}

\begin{itemize}

\item {\textbf{States of specific families}}\\
First, in Table \ref{tab:examples} we give the optimal parameters found for the classes of states studied, along with the amount of entanglement in the optimal example. In each case we studied 6 different sets of measurements, such that the measurement directions are aligned along the vertices of different regular solids.

\begin{table*}
    \begin{tabular}{ l|l|c|c|c|c|c}
        \multicolumn{1}{c|}{Solid} & \multicolumn{1}{c|}{(Vertices,} & \multicolumn{2}{c|}{Werner states} & \multicolumn{2}{c|}{ Bell diagonal states} & Witnessed \\
         & \multicolumn{1}{c|}{Radius)} & \multicolumn{2}{c|}{$\rho_\mathrm{w}(w)$} & \multicolumn{2}{c|}{$\rho_\mathrm{bell}(p_1,p_2,p_3)$ } & states \\ \hline
         & & $w^*$ & $\mathcal{N}$ & $p_1^*$ & $\mathcal{N}$ & $\mathcal{N}$ \\ \cline{3-7}
        Icosahedron & (12, 0.79) & 0.4285 & 0.0714  & 0.5390 & 0.0390 & 0.0754 \\ \hline
        Dodecahedron & (20, 0.79) & 0.4160 & 0.0620  & 0.5296 & 0.0296 & 0.0647 \\ \hline
        Truncated cube & (24, 0.67) & 0.3553 & 0.0164  & 0.500 & 0 & 0.0181 \\ \hline
        Truncated octahedron & (24, 0.77) & 0.4082 & 0.0561  & 0.5071 & 0.0071 & 0.0601 \\ \hline
        Truncated tetrahedron + antipodal\footnote{Antipodal points are added because the vertices of the truncated tetrahedron are not symmetric with respect to the origin.} & (24, 0.85) & 0.4404 & 0.0803  & 0.5581 & 0.0581 & 0.0839 \\ \hline
        Rhombicuboctahedron & (24, 0.86) & 0.4454 & 0.0840  & 0.5664 & 0.0664 & 0.0883                               
    \end{tabular}
    \caption{\label{tab:examples} Examples of optimal entangled LHS states within specific families -- Werner states, rank 3 Bell diagonal states, and witnessed states -- with measurements performed along the vertices of Platonic and Archimedian solids. The table presents the values of the optimized parameters of merit and the negativity $\mathcal{N}$ of the corresponding state. The entanglement witness considered for the witnessed state is the one-sided generalised robustness}
\end{table*}

\item \textbf{States with LHS models for projective measurements, obtained with uniformly chosen witnesses and 6 uniformly chosen measurements on the Bloch sphere.}\\

We collected 450000 such states. In Fig.~\ref{fig:hist_normalized_examples_Paul_Proj_Rand6}, we present a normalized histogram of their negativities, and in Fig.~\ref{fig:trace_normalized_sample_9000_examples_Paul_Proj_Rand6} we present a normalized histogram of trace distances between every two states in a random sample of 9000 such states. The mean negativity of all 450000 states is $\expect{E_{\textrm{Neg}}} = 0.016$, with standard deviation $\sigma_{\textrm{Neg}} = 0.008$. The mean trace distance of the states in the sample is $\expect{T} = 0.602$, with standard deviation $\sigma_{T} = 0.124$.

\begin{figure*}[h!]
\centering
	\begin{subfigure}[b]{0.4\textwidth}
		\includegraphics[width=\textwidth]{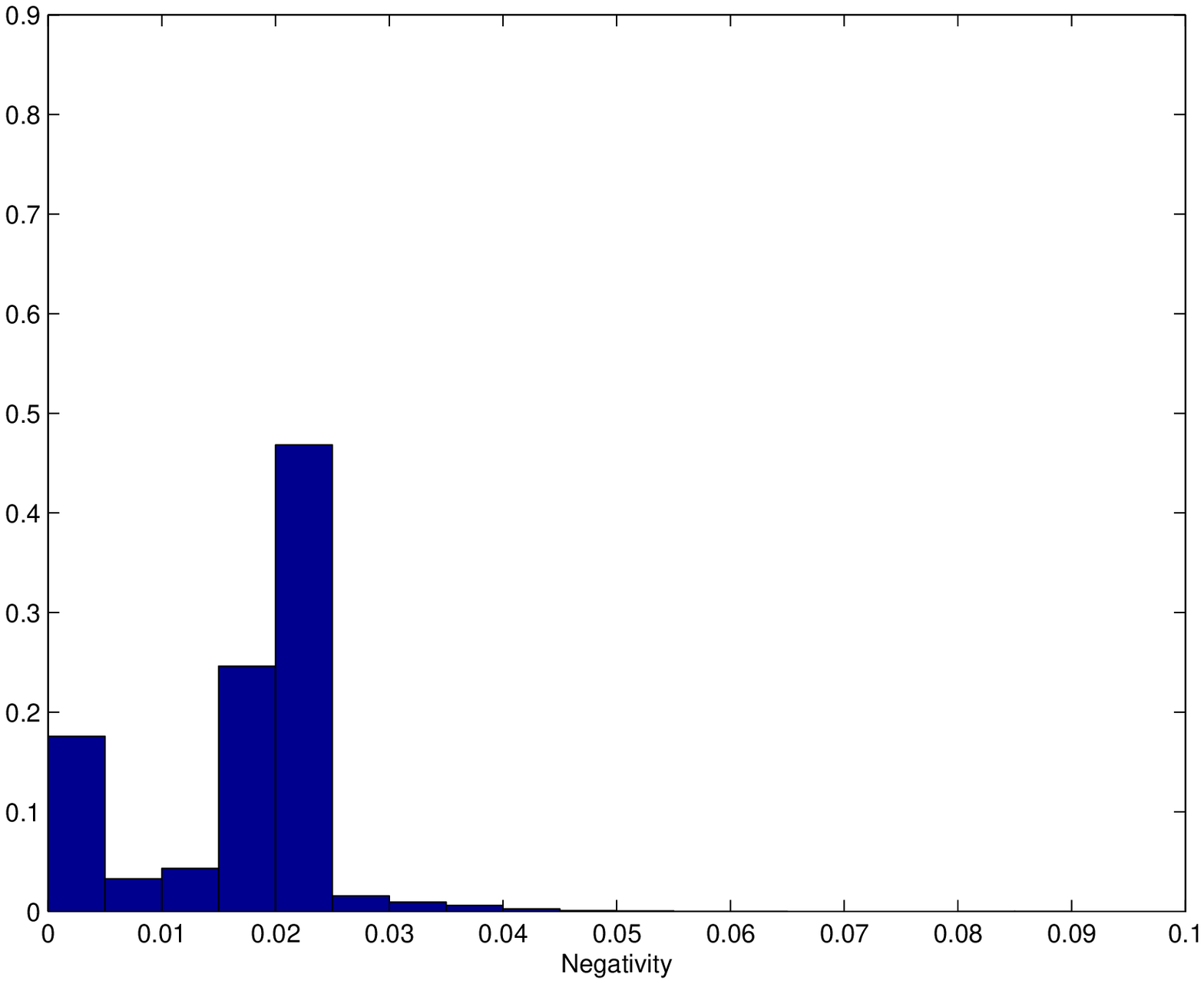}
		\caption{Negativity histogram}
		\label{fig:hist_normalized_examples_Paul_Proj_Rand6}
	\end{subfigure}
	\begin{subfigure}[b]{0.4\textwidth}
		\includegraphics[width=\textwidth]{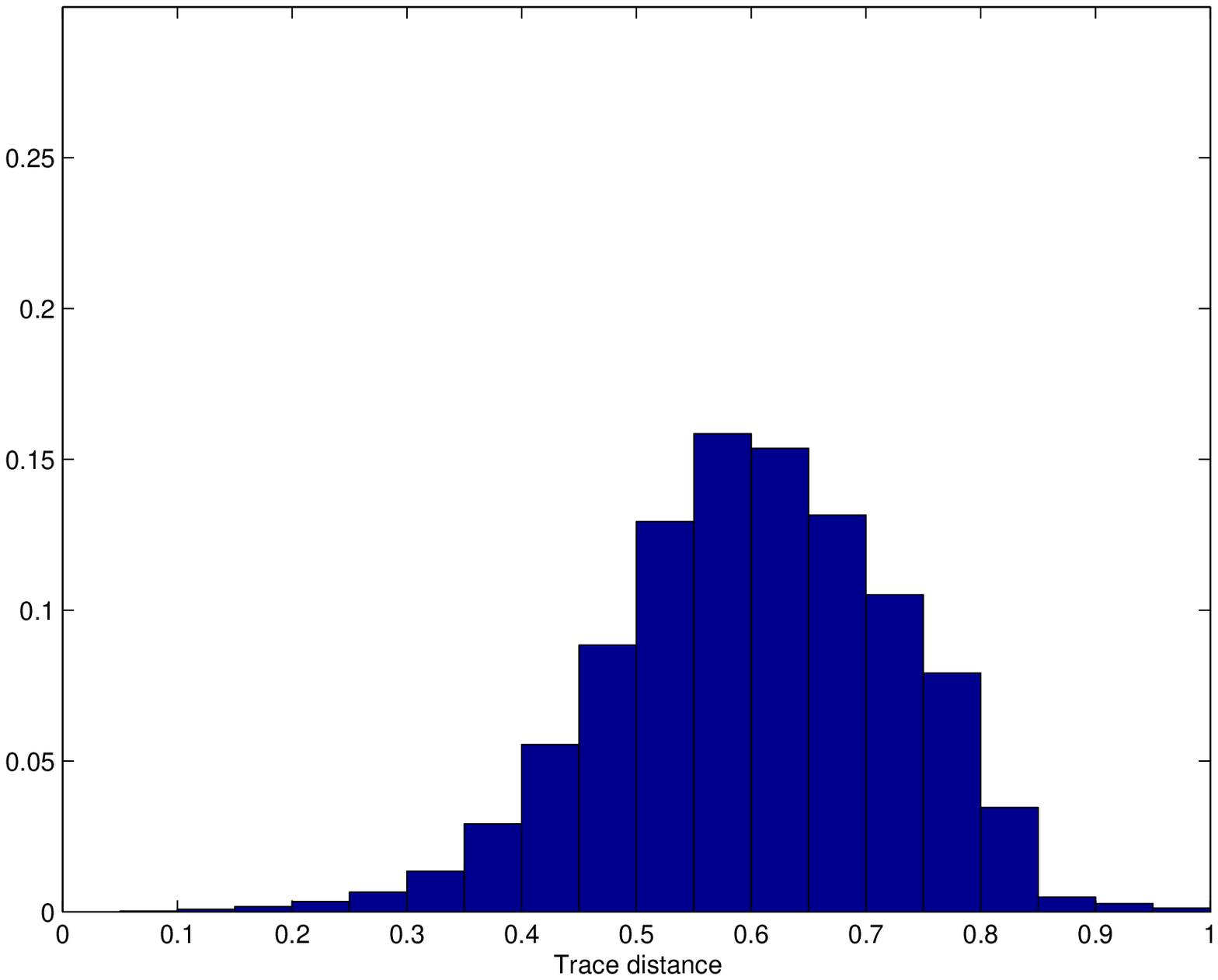}
		\caption{Trace distance histogram}		\label{fig:trace_normalized_sample_9000_examples_Paul_Proj_Rand6}
	\end{subfigure}
	\caption{Normalized histograms of the states with LHS models for projective measurements, obtained with uniformly chosen entanglement witnesses and 6 uniformly chosen measurements on the Bloch sphere.}
\end{figure*}

\item \textbf{States with LHS models for projective measurements, obtained with uniformly chosen witnesses and measurements along the vertices of the icosahedron.}\\
We collected 300000 such states. In Fig.~\ref{fig:hist_normalized_examples_proj_ico}, we present a normalized histogram of their negativities, and in Fig.~\ref{fig:trace_normalized_sample_6000_examples_proj_ico} we present a normalized histogram of trace distances between every two states in a random sample of 6000 such states. The mean negativity of all 300000 states is $\expect{E_{\textrm{Neg}}} = 0.066$, with standard deviation $\sigma_{\textrm{Neg}} = 0.016$. The mean trace distance of the states in the sample is $\expect{T} = 0.549$, with standard deviation $\sigma_{T} = 0.103$.

\begin{figure*}[h!]
\centering
	\begin{subfigure}[h]{0.4\textwidth}
		\includegraphics[width=\textwidth]{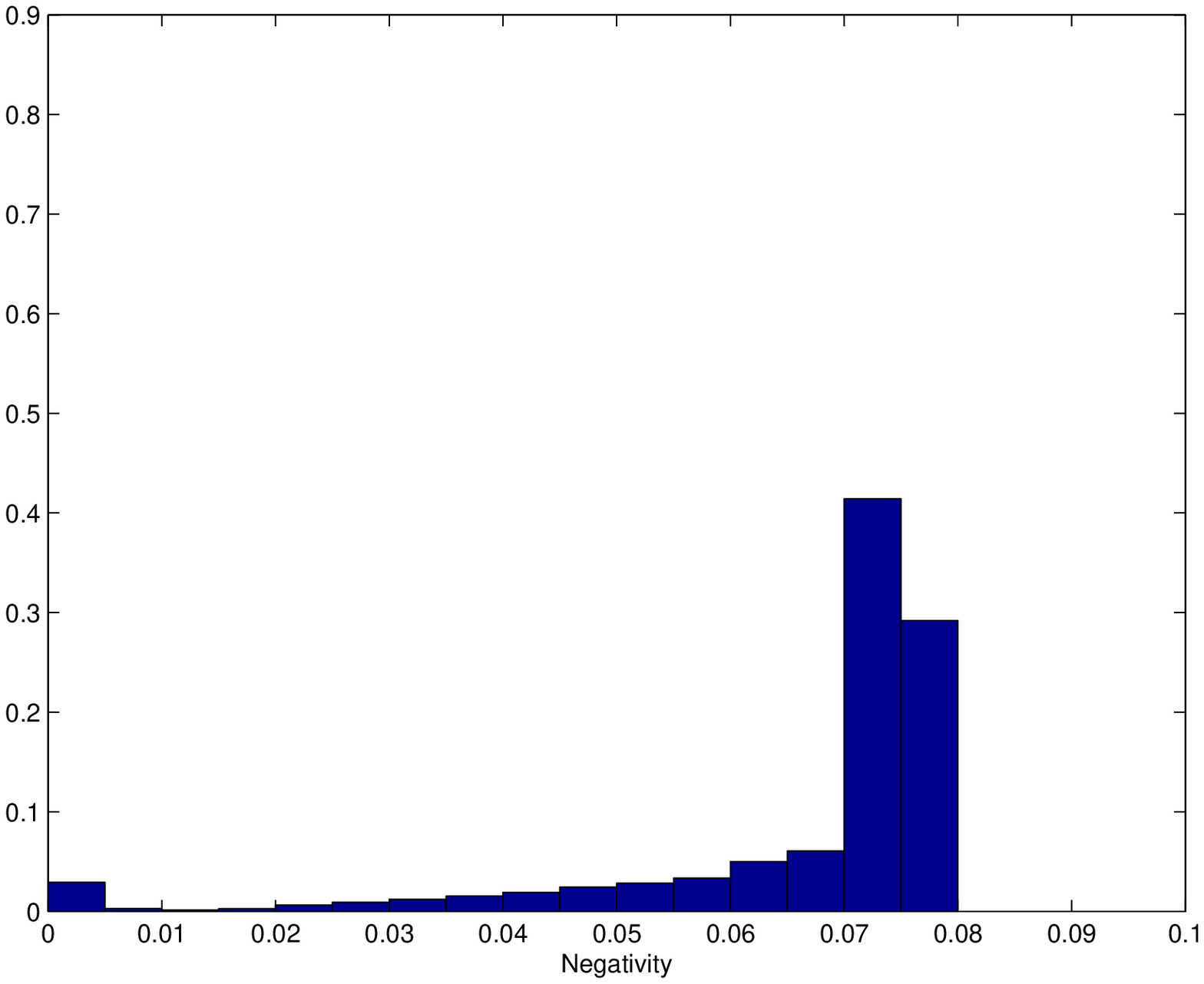}
		\caption{Negativity histogram}
		\label{fig:hist_normalized_examples_proj_ico}
	\end{subfigure}
	\begin{subfigure}[h]{0.4\textwidth}
		\includegraphics[width=\textwidth]{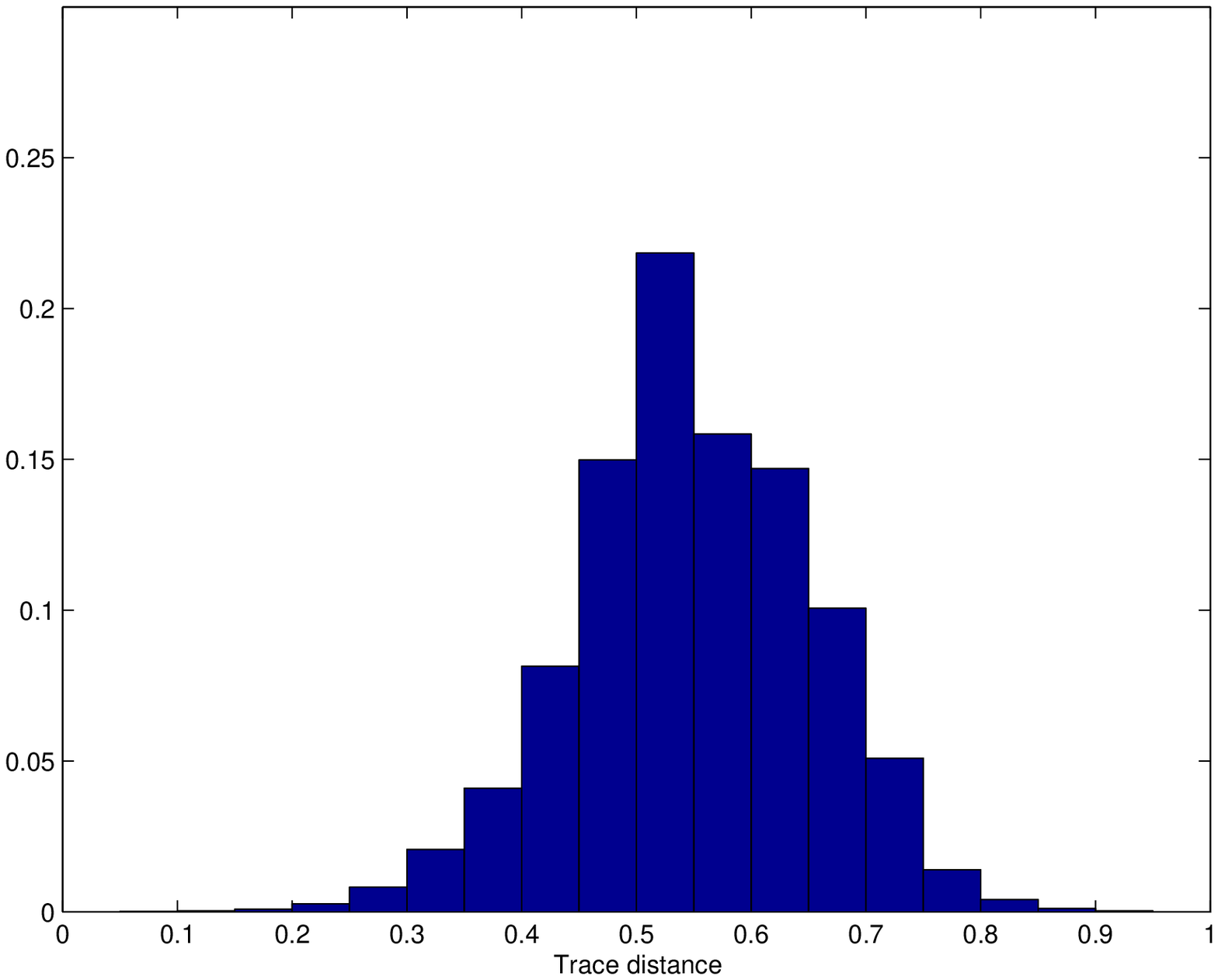}
		\caption{Trace distance histogram}
		\label{fig:trace_normalized_sample_6000_examples_proj_ico}
	\end{subfigure}
	\caption{Normalized histograms for the states with LHS models for projective measurements, obtained with uniformly chosen witnesses and measurements along the vertices of the icosahedron.}
\end{figure*}

\item \textbf{States with LHS models for projective measurements, obtained from uniformly chosen two-qubit density matrices according to the Hilbert-Schmidt measure.}\\
Of $20000$ two-qubit states we drawed according to the Hilbert-Schmidt measure, we obtained $15228$ entangled states $(\approx 76\%)$, certified by means of the Peres-Horodecki criterion. Among all entangled states, we were able to certify that $2961$ states $(\approx 19\%)$ admit LHS models for projective measurements by performing the SDP test we present in the main text, considering projective measurements along the vertices of the icosahedron.

In Fig.~\ref{fig:hist_normalized_examples_HS_proj_ico} we present a  normalized histogram of the of the $2961$ entangled states with LHS models we obtained, and in Fig.~\ref{fig:trace_normalized_examples_HS_proj_ico} we present a normalized histogram of trace distances between every two states such states. The mean negativity of all $2961$ states is $\expect{E_{\textrm{Neg}}} = 0.020$, with standard deviation $\sigma_{\textrm{Neg}} = 0.013$. The mean trace distance of the states is $\expect{T} = 0.520$, with standard deviation $\sigma_{T} = 0.083$.

\begin{figure*}[h!]
\centering
	\begin{subfigure}[h]{0.4\textwidth}
		\includegraphics[width=\textwidth]{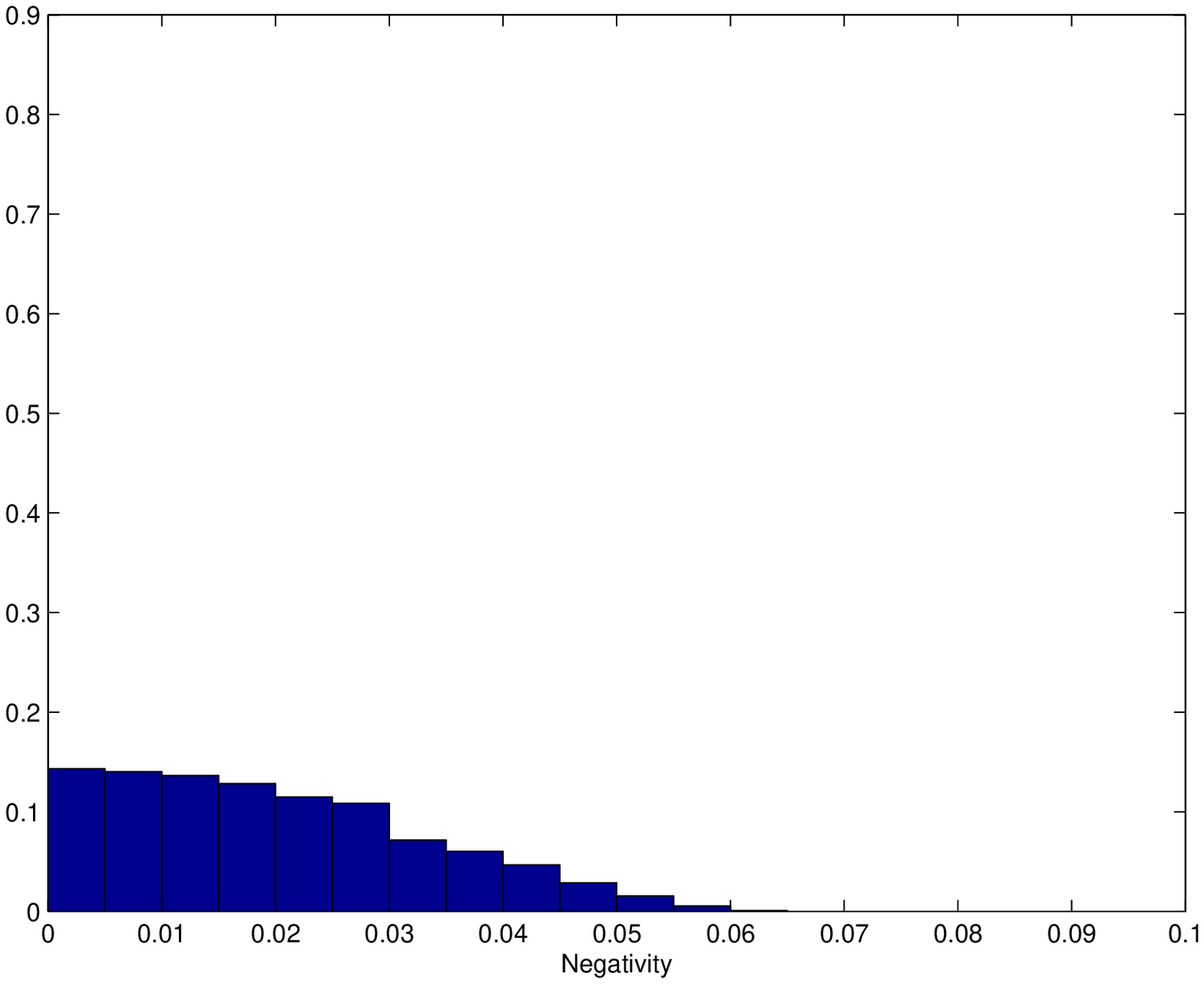}
		\caption{Negativity histogram}
		\label{fig:hist_normalized_examples_HS_proj_ico}
	\end{subfigure}
	\begin{subfigure}[h]{0.4\textwidth}
		\includegraphics[width=\textwidth]{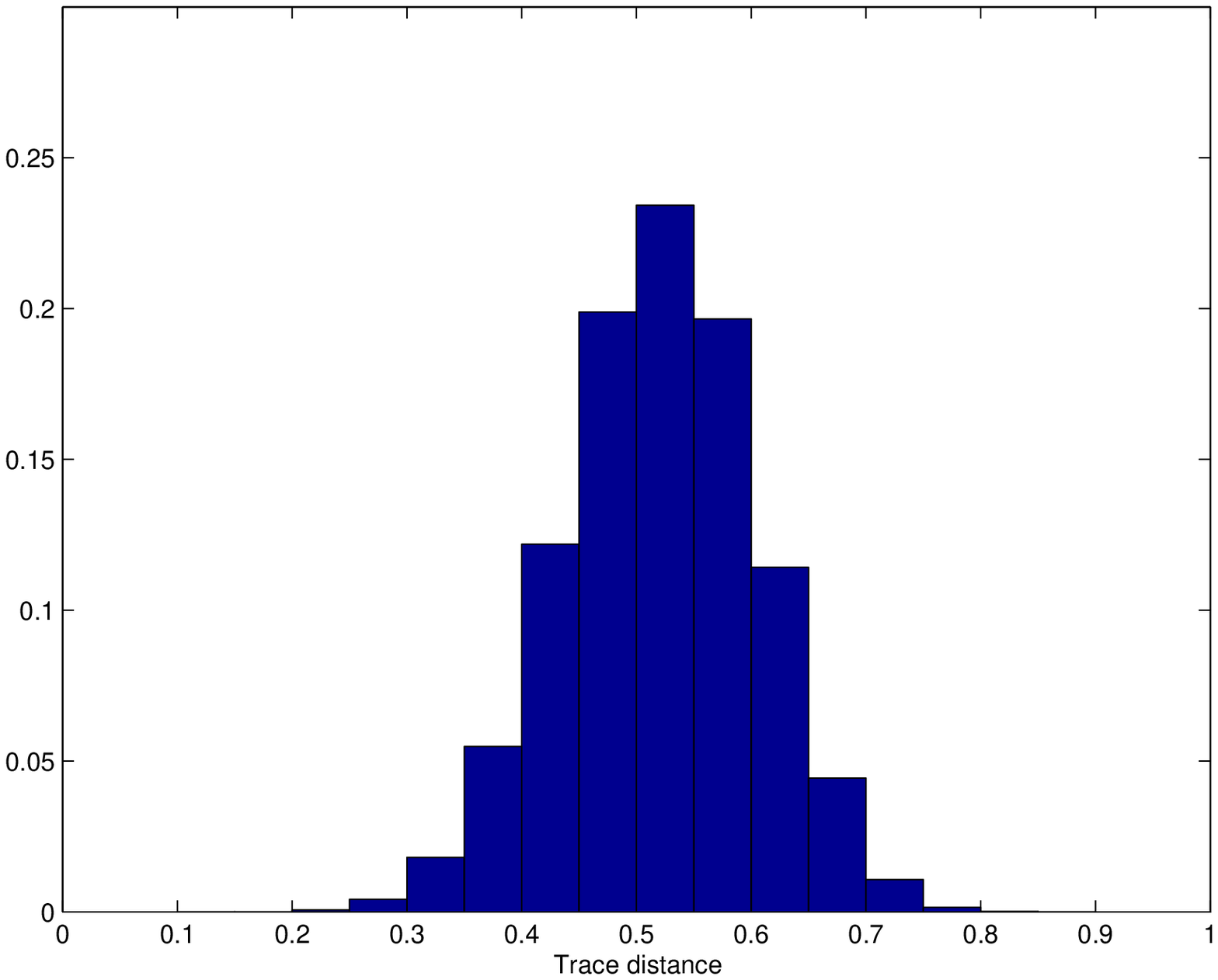}
		\caption{Trace distance histogram}
		\label{fig:trace_normalized_examples_HS_proj_ico}
	\end{subfigure}
	\caption{Normalized histograms for the states with LHS models for projective measurements, obtained from uniformly chosen two-qubit density matrices according to the Hilbert-Schmidt measure.}
\end{figure*}

\item \textbf{States with LHS models for projective measurements, obtained from uniformly chosen two-qubit density matrices according to the Bures measure.}\\
Of $20000$ two-qubit states we drew according to the Bures measure, we obtained $18447$ entangled states $(\approx 92\%)$, certified by means of the Peres-Horodecki criterion. Among all entangled states, we were able to certify that $932$ states $(\approx 5\%)$ admit LHS models for projective measurements by performing the SDP test we present in the main text, considering projective measurements along the vertices of the icosahedron.

In Fig.~\ref{fig:hist_normalized_examples_bures_proj_ico} we present a  normalized histogram of the of the $932$ entangled states with LHS models we obtained, and in Fig.~\ref{fig:trace_normalized_examples_bures_proj_ico} we present a normalized histogram of trace distances between every two such states. The mean negativity of all $932$ states is $\expect{E_{\textrm{Neg}}} = 0.018$, with standard deviation $\sigma_{\textrm{Neg}} = 0.012$. The mean trace distance of the states is $\expect{T} = 0.556$, with standard deviation $\sigma_{T} = 0.090$.

\begin{figure*}[h!]
\centering
	\begin{subfigure}[h]{0.4\textwidth}
		\includegraphics[width=\textwidth]{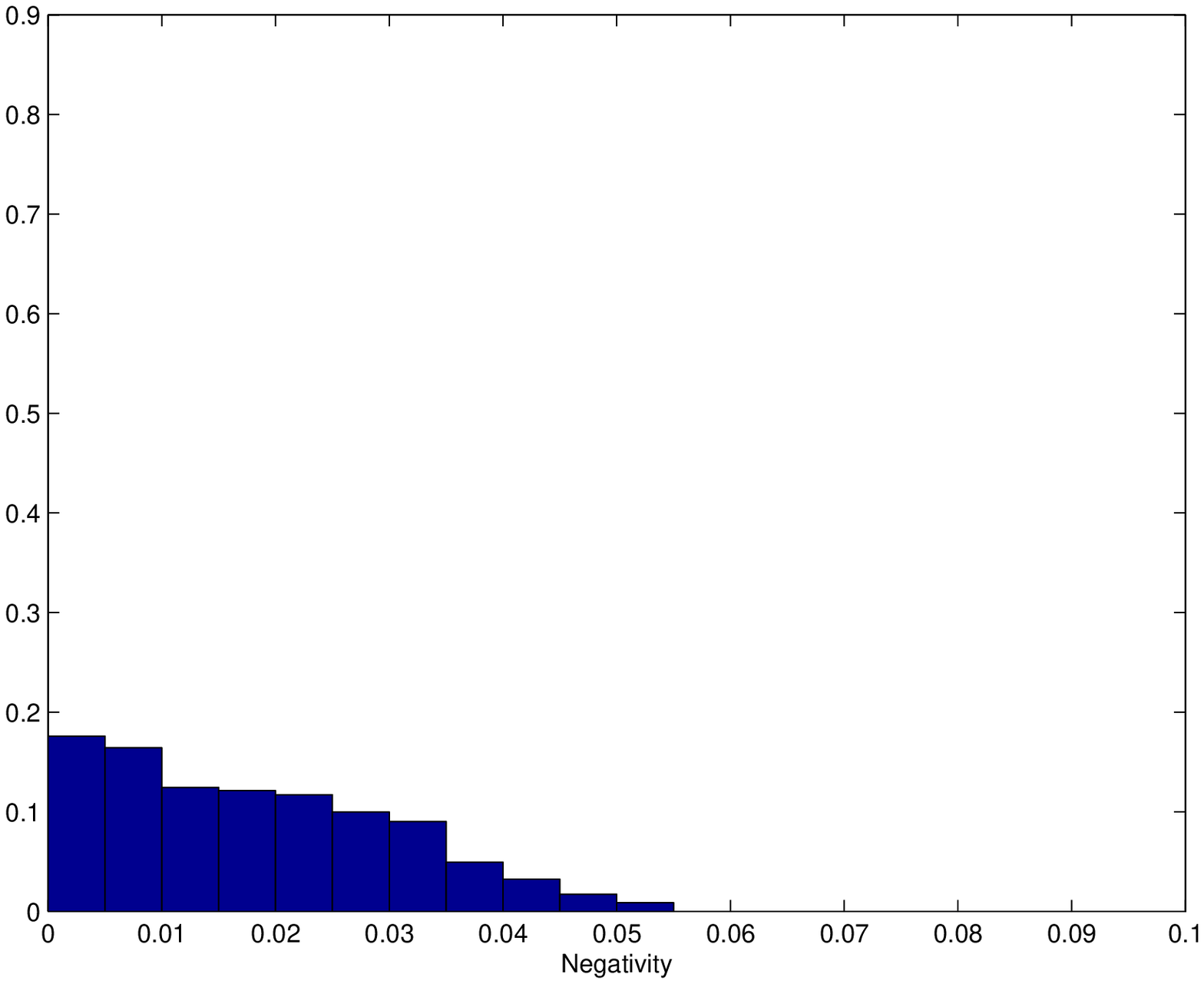}
		\caption{Negativity histogram}
		\label{fig:hist_normalized_examples_bures_proj_ico}
	\end{subfigure}
	\begin{subfigure}[h]{0.4\textwidth}
		\includegraphics[width=\textwidth]{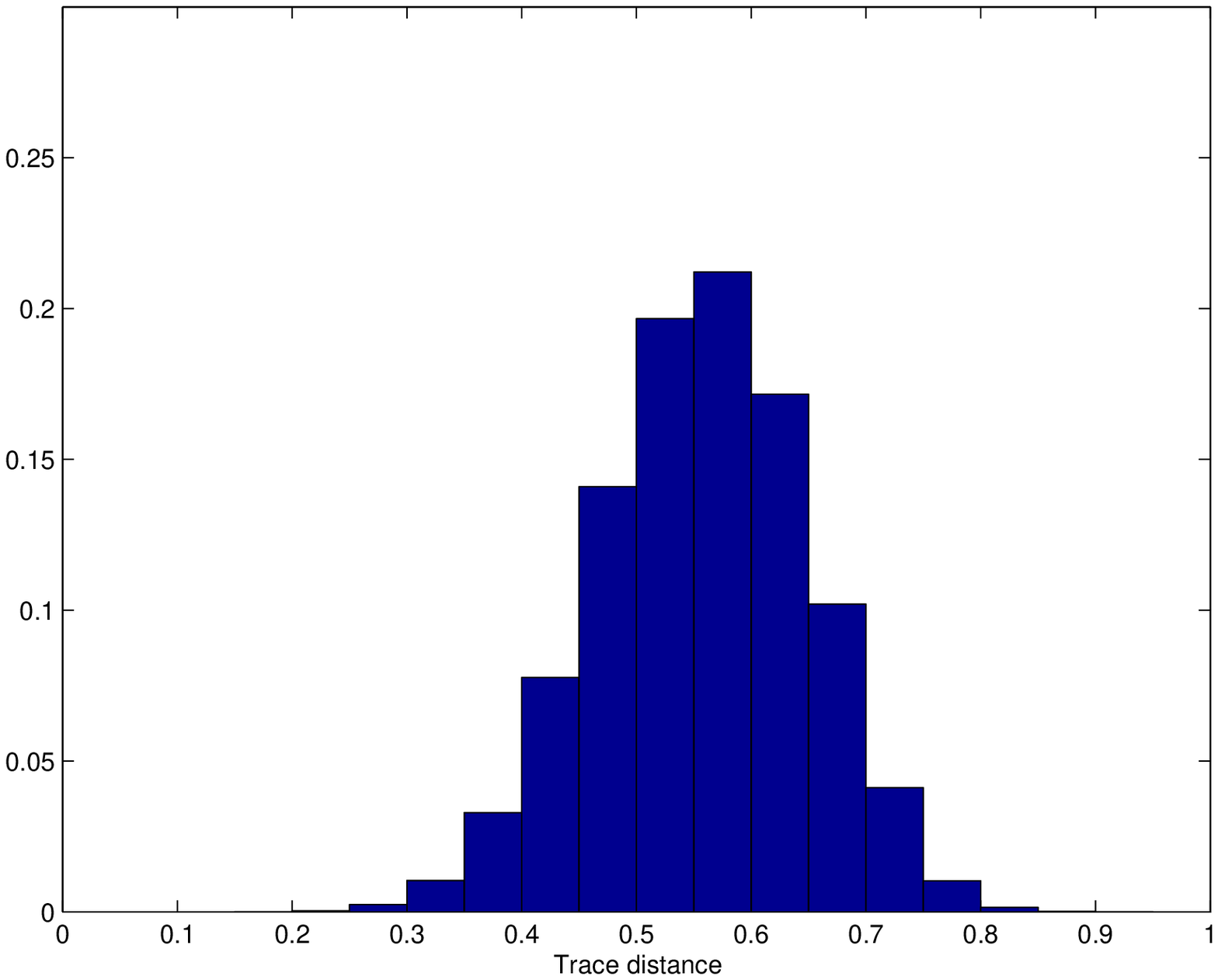}
		\caption{Trace distance histogram}
		\label{fig:trace_normalized_examples_bures_proj_ico}
	\end{subfigure}
	\caption{Normalized histograms for the states with LHS models for projective measurements, obtained from uniformly chosen two-qubit density matrices according to the Bures measure.}
\end{figure*}

\item \textbf{States with LHS models for POVMs, obtained with uniformly chosen witnesses and measurements along the vertices of the rhombicuboctahedron.}\\
We collected 1500 such states. In Fig.~\ref{fig:hist_normalized_examples_paul_geneva_rhombi_1500} we present a normalized histogram of their negativities, and in Fig.~\ref{fig:trace_normalized_examples_paul_geneva_rhombi_1500} we present a normalized histogram of trace distances between every two states. The mean negativity of all 1500 states is $\expect{E_{\textrm{Neg}}} = 0.008$, with standard deviation $\sigma_{\textrm{Neg}} = 0.002$. The mean trace distance of the states is $\expect{T} = 0.496$, with standard deviation $\sigma_{T} = 0.175$.

\begin{figure*}[h!]
\centering
	\begin{subfigure}[h]{0.4\textwidth}
		\includegraphics[width=\textwidth]{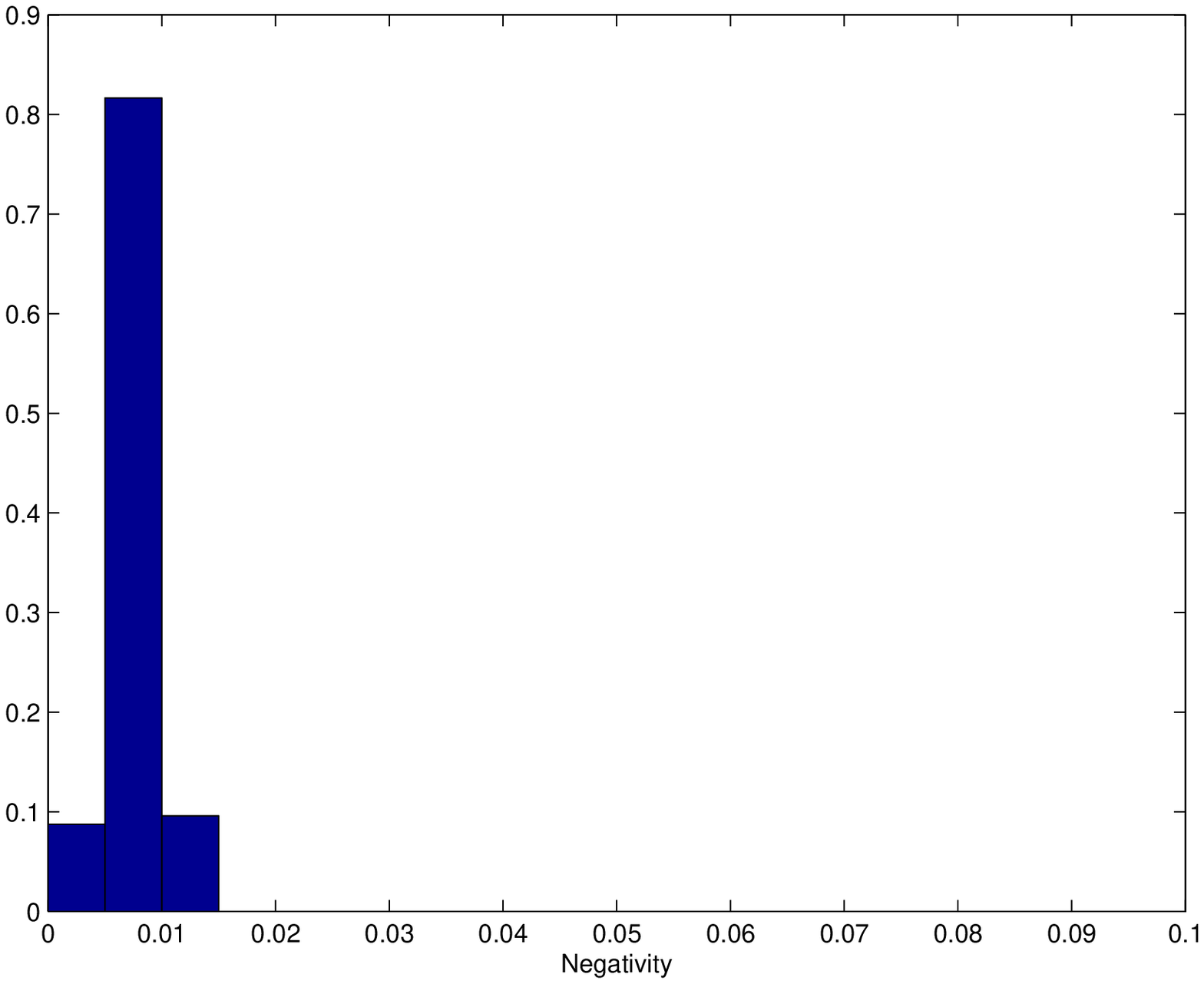}
		\caption{Negativity histogram}
		\label{fig:hist_normalized_examples_paul_geneva_rhombi_1500}
	\end{subfigure}
	\begin{subfigure}[h]{0.4\textwidth}
		\includegraphics[width=\textwidth]{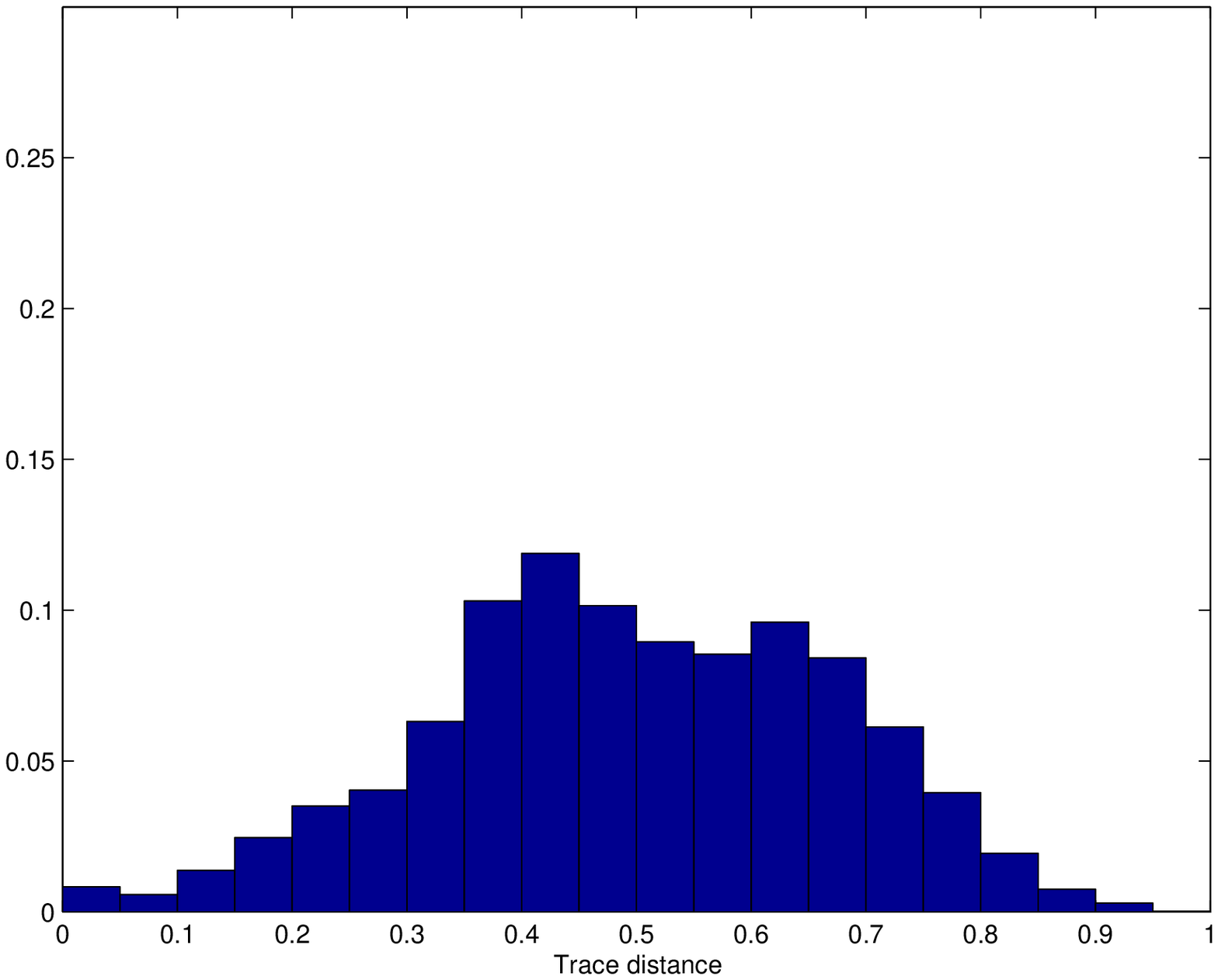}
		\caption{Trace distance histogram}
		\label{fig:trace_normalized_examples_paul_geneva_rhombi_1500}
	\end{subfigure}
	\caption{Normalized histograms for the states with LHS models for POVMs, obtained with uniformly chosen witnesses and measurements along the vertices of the rhombicuboctahedron.}
\end{figure*}

\end{itemize}

\section{Appendix 3. Genuinely multipartite entangled states with LHS models}

By considering a class of entanglement witnesses that are sensitive only to genuine multipartite entanglement, we adapted our second method and were able to find examples of genuine tripartite entangled states that admit LHS models.

In \cite{guhne}, it was shown that a given state $\rho$ can be proven to be genuinely multipartite entangled if there exists a witness $W$ such that $\tr[{W\rho}] < 0$ and $W$ is \emph{fully decomposable}, that is, there exists positive semidefinite operators $P_{M}$ and $Q_{M}$ such that $W = P_{M} + Q_{M}^{T_{M}}$, for all partitions $M$ of the system, where $T_{M}$ denotes partial transposition with respect to $M$. This test can be implemented as the following SDP:
\begin{align}
 \min_{W} \quad & \tr[W\rho]  \nonumber  \\
 \mathrm{s.t.} \quad &\tr[W] = 1, \nonumber  \\
  & W = P_{M} + Q_{M}^{T_{M}} \,\, \forall M,  \\
  & P_{M} \geq 0 \,\, \forall M, \quad Q_{M} \geq 0\,\, \forall M. \nonumber
\end{align}

The search for genuine tripartite entangled states with LHS models followed the same method used to search for random entangled states with LHS models. We start with a random pure three-qubit state, and use the above SDP to find the best witness $W$ that detects its genuine tripartite entanglement. Fixing, now, the witness $W$, we apply the SDP of method 2 to obtain the LHS state that minimizes the expected value of $W$ -- the collection of measurements $\mathcal{M}$ we chose are the projective measurements along the vertices of the rhombicuboctahedron. If $\tr({W\rho_{\rA\rB\rC}}) < 0$, we obtain a genuine tripartite entangled state with a LHS model for all projective measurements. Analogously to method 2, we iterate the process, obtaining a better witness based on $\rho_{\rA\rB\rC}$, and so on. By this procedure we were able to obtain 150 genuine tripartite entangled states with LHS models for all projective measurements. We were not able, though, to find any genuine tripartite entangled state with LHS model for POVMs.

\end{appendix}


\begin{thebibliography}{35}%
\makeatletter
\providecommand \@ifxundefined [1]{%
 \@ifx{#1\undefined}
}%
\providecommand \@ifnum [1]{%
 \ifnum #1\expandafter \@firstoftwo
 \else \expandafter \@secondoftwo
 \fi
}%
\providecommand \@ifx [1]{%
 \ifx #1\expandafter \@firstoftwo
 \else \expandafter \@secondoftwo
 \fi
}%
\providecommand \natexlab [1]{#1}%
\providecommand \enquote  [1]{``#1''}%
\providecommand \bibnamefont  [1]{#1}%
\providecommand \bibfnamefont [1]{#1}%
\providecommand \citenamefont [1]{#1}%
\providecommand \href@noop [0]{\@secondoftwo}%
\providecommand \href [0]{\begingroup \@sanitize@url \@href}%
\providecommand \@href[1]{\@@startlink{#1}\@@href}%
\providecommand \@@href[1]{\endgroup#1\@@endlink}%
\providecommand \@sanitize@url [0]{\catcode `\\12\catcode `\$12\catcode
  `\&12\catcode `\#12\catcode `\^12\catcode `\_12\catcode `\%12\relax}%
\providecommand \@@startlink[1]{}%
\providecommand \@@endlink[0]{}%
\providecommand \url  [0]{\begingroup\@sanitize@url \@url }%
\providecommand \@url [1]{\endgroup\@href {#1}{\urlprefix }}%
\providecommand \urlprefix  [0]{URL }%
\providecommand \Eprint [0]{\href }%
\providecommand \doibase [0]{http://dx.doi.org/}%
\providecommand \selectlanguage [0]{\@gobble}%
\providecommand \bibinfo  [0]{\@secondoftwo}%
\providecommand \bibfield  [0]{\@secondoftwo}%
\providecommand \translation [1]{[#1]}%
\providecommand \BibitemOpen [0]{}%
\providecommand \bibitemStop [0]{}%
\providecommand \bibitemNoStop [0]{.\EOS\space}%
\providecommand \EOS [0]{\spacefactor3000\relax}%
\providecommand \BibitemShut  [1]{\csname bibitem#1\endcsname}%
\let\auto@bib@innerbib\@empty
\bibitem [{\citenamefont {Brunner}\ \emph {et~al.}(2014)\citenamefont
  {Brunner}, \citenamefont {Cavalcanti}, \citenamefont {Pironio}, \citenamefont
  {Scarani},\ and\ \citenamefont {Wehner}}]{BruCavPir14}%
  \BibitemOpen
  \bibfield  {author} {\bibinfo {author} {\bibfnamefont {N.}~\bibnamefont
  {Brunner}}, \bibinfo {author} {\bibfnamefont {D.}~\bibnamefont {Cavalcanti}},
  \bibinfo {author} {\bibfnamefont {S.}~\bibnamefont {Pironio}}, \bibinfo
  {author} {\bibfnamefont {V.}~\bibnamefont {Scarani}}, \ and\ \bibinfo
  {author} {\bibfnamefont {S.}~\bibnamefont {Wehner}},\ }\href {\doibase
  10.1103/RevModPhys.86.419} {\bibfield  {journal} {\bibinfo  {journal} {Rev.
  Mod. Phys.}\ }\textbf {\bibinfo {volume} {86}},\ \bibinfo {pages} {419}
  (\bibinfo {year} {2014})}\BibitemShut {NoStop}%
\bibitem [{\citenamefont {Bell}(1964)}]{Bel64}%
  \BibitemOpen
  \bibfield  {author} {\bibinfo {author} {\bibfnamefont {J.~S.}\ \bibnamefont
  {Bell}},\ }\href@noop {} {\bibfield  {journal} {\bibinfo  {journal}
  {Physics}\ }\textbf {\bibinfo {volume} {1}},\ \bibinfo {pages} {195}
  (\bibinfo {year} {1964})}\BibitemShut {NoStop}%
\bibitem [{\citenamefont {Werner}(1989)}]{Werner_1989}%
  \BibitemOpen
  \bibfield  {author} {\bibinfo {author} {\bibfnamefont {R.~F.}\ \bibnamefont
  {Werner}},\ }\href {\doibase 10.1103/physreva.40.4277} {\bibfield  {journal}
  {\bibinfo  {journal} {Phys. Rev. A}\ }\textbf {\bibinfo {volume} {40}},\
  \bibinfo {pages} {4277} (\bibinfo {year} {1989})}\BibitemShut {NoStop}%
\bibitem [{\citenamefont {Barrett}(2002)}]{Barrett_2002}%
  \BibitemOpen
  \bibfield  {author} {\bibinfo {author} {\bibfnamefont {J.}~\bibnamefont
  {Barrett}},\ }\href {\doibase 10.1103/physreva.65.042302} {\bibfield
  {journal} {\bibinfo  {journal} {Phys. Rev. A}\ }\textbf {\bibinfo {volume}
  {65}} (\bibinfo {year} {2002}),\ 10.1103/physreva.65.042302}\BibitemShut
  {NoStop}%
\bibitem [{\citenamefont {Ac{\'{\i}}n}\ \emph {et~al.}(2006)\citenamefont
  {Ac{\'{\i}}n}, \citenamefont {Gisin},\ and\ \citenamefont
  {Toner}}]{Ac_n_2006}%
  \BibitemOpen
  \bibfield  {author} {\bibinfo {author} {\bibfnamefont {A.}~\bibnamefont
  {Ac{\'{\i}}n}}, \bibinfo {author} {\bibfnamefont {N.}~\bibnamefont {Gisin}},
  \ and\ \bibinfo {author} {\bibfnamefont {B.}~\bibnamefont {Toner}},\ }\href
  {\doibase 10.1103/physreva.73.062105} {\bibfield  {journal} {\bibinfo
  {journal} {Phys. Rev. A}\ }\textbf {\bibinfo {volume} {73}} (\bibinfo {year}
  {2006}),\ 10.1103/physreva.73.062105}\BibitemShut {NoStop}%
\bibitem [{\citenamefont {Almeida}\ \emph {et~al.}(2007)\citenamefont
  {Almeida}, \citenamefont {Pironio}, \citenamefont {Barrett}, \citenamefont
  {T{\'{o}}th},\ and\ \citenamefont {Ac{\'{\i}}n}}]{Almeida_2007}%
  \BibitemOpen
  \bibfield  {author} {\bibinfo {author} {\bibfnamefont {M.~L.}\ \bibnamefont
  {Almeida}}, \bibinfo {author} {\bibfnamefont {P.}~\bibnamefont {Pironio}},
  \bibinfo {author} {\bibfnamefont {J.}~\bibnamefont {Barrett}}, \bibinfo
  {author} {\bibfnamefont {G.}~\bibnamefont {T{\'{o}}th}}, \ and\ \bibinfo
  {author} {\bibfnamefont {A.}~\bibnamefont {Ac{\'{\i}}n}},\ }\href {\doibase
  10.1103/physrevlett.99.040403} {\bibfield  {journal} {\bibinfo  {journal}
  {Phys. Rev. Lett.}\ }\textbf {\bibinfo {volume} {99}} (\bibinfo {year}
  {2007}),\ 10.1103/physrevlett.99.040403}\BibitemShut {NoStop}%
\bibitem [{\citenamefont {Wiseman}\ \emph {et~al.}(2007)\citenamefont
  {Wiseman}, \citenamefont {Jones},\ and\ \citenamefont
  {Doherty}}]{Wiseman_2007}%
  \BibitemOpen
  \bibfield  {author} {\bibinfo {author} {\bibfnamefont {H.~M.}\ \bibnamefont
  {Wiseman}}, \bibinfo {author} {\bibfnamefont {S.~J.}\ \bibnamefont {Jones}},
  \ and\ \bibinfo {author} {\bibfnamefont {A.~C.}\ \bibnamefont {Doherty}},\
  }\href {\doibase 10.1103/physrevlett.98.140402} {\bibfield  {journal}
  {\bibinfo  {journal} {Phys. Rev. Lett.}\ }\textbf {\bibinfo {volume} {98}}
  (\bibinfo {year} {2007}),\ 10.1103/physrevlett.98.140402}\BibitemShut
  {NoStop}%
\bibitem [{\citenamefont {Jevtic}\ \emph {et~al.}(2015)\citenamefont {Jevtic},
  \citenamefont {Hall}, \citenamefont {Anderson}, \citenamefont {Zwierz},\ and\
  \citenamefont {Wiseman}}]{Jevtic_2015}%
  \BibitemOpen
  \bibfield  {author} {\bibinfo {author} {\bibfnamefont {S.}~\bibnamefont
  {Jevtic}}, \bibinfo {author} {\bibfnamefont {M.~J.~W.}\ \bibnamefont {Hall}},
  \bibinfo {author} {\bibfnamefont {M.~R.}\ \bibnamefont {Anderson}}, \bibinfo
  {author} {\bibfnamefont {M.}~\bibnamefont {Zwierz}}, \ and\ \bibinfo {author}
  {\bibfnamefont {H.~M.}\ \bibnamefont {Wiseman}},\ }\href {\doibase
  10.1364/josab.32.000a40} {\bibfield  {journal} {\bibinfo  {journal} {J. Opt.
  Soc. Am. B}\ }\textbf {\bibinfo {volume} {32}},\ \bibinfo {pages} {A40}
  (\bibinfo {year} {2015})}\BibitemShut {NoStop}%
\bibitem [{\citenamefont {Hirsch}\ \emph {et~al.}(2013)\citenamefont {Hirsch},
  \citenamefont {Quintino}, \citenamefont {Bowles},\ and\ \citenamefont
  {Brunner}}]{Hirsch_2013}%
  \BibitemOpen
  \bibfield  {author} {\bibinfo {author} {\bibfnamefont {F.}~\bibnamefont
  {Hirsch}}, \bibinfo {author} {\bibfnamefont {M.~T.}\ \bibnamefont
  {Quintino}}, \bibinfo {author} {\bibfnamefont {J.}~\bibnamefont {Bowles}}, \
  and\ \bibinfo {author} {\bibfnamefont {N.}~\bibnamefont {Brunner}},\ }\href
  {\doibase 10.1103/physrevlett.111.160402} {\bibfield  {journal} {\bibinfo
  {journal} {Phys. Rev. Lett.}\ }\textbf {\bibinfo {volume} {111}} (\bibinfo
  {year} {2013}),\ 10.1103/physrevlett.111.160402}\BibitemShut {NoStop}%
\bibitem [{\citenamefont {Augusiak}\ \emph {et~al.}(2014)\citenamefont
  {Augusiak}, \citenamefont {Demianowicz},\ and\ \citenamefont
  {Ac{\'{\i}}n}}]{Augusiak_2014}%
  \BibitemOpen
  \bibfield  {author} {\bibinfo {author} {\bibfnamefont {R.}~\bibnamefont
  {Augusiak}}, \bibinfo {author} {\bibfnamefont {M.}~\bibnamefont
  {Demianowicz}}, \ and\ \bibinfo {author} {\bibfnamefont {A.}~\bibnamefont
  {Ac{\'{\i}}n}},\ }\href {\doibase 10.1088/1751-8113/47/42/424002} {\bibfield
  {journal} {\bibinfo  {journal} {J. Phys. A: Math.Theor.}\ }\textbf {\bibinfo
  {volume} {47}},\ \bibinfo {pages} {424002} (\bibinfo {year}
  {2014})}\BibitemShut {NoStop}%
\bibitem [{\citenamefont {Bowles}\ \emph
  {et~al.}(2016{\natexlab{a}})\citenamefont {Bowles}, \citenamefont {Hirsch},
  \citenamefont {Quintino},\ and\ \citenamefont {Brunner}}]{Bowles2015}%
  \BibitemOpen
  \bibfield  {author} {\bibinfo {author} {\bibfnamefont {J.}~\bibnamefont
  {Bowles}}, \bibinfo {author} {\bibfnamefont {F.}~\bibnamefont {Hirsch}},
  \bibinfo {author} {\bibfnamefont {M.~T.}\ \bibnamefont {Quintino}}, \ and\
  \bibinfo {author} {\bibfnamefont {N.}~\bibnamefont {Brunner}},\ }\href
  {\doibase 10.1103/PhysRevA.93.022121} {\bibfield  {journal} {\bibinfo
  {journal} {Phys. Rev. A}\ }\textbf {\bibinfo {volume} {93}},\ \bibinfo
  {pages} {022121} (\bibinfo {year} {2016}{\natexlab{a}})}\BibitemShut
  {NoStop}%
\bibitem [{\citenamefont {Schr{\"{o}}dinger}(1936)}]{Schr_dinger_1936}%
  \BibitemOpen
  \bibfield  {author} {\bibinfo {author} {\bibfnamefont {E.}~\bibnamefont
  {Schr{\"{o}}dinger}},\ }\href {\doibase 10.1017/s0305004100019137} {\bibfield
   {journal} {\bibinfo  {journal} {Math. Proc. Camb. Phil. Soc.}\ }\textbf
  {\bibinfo {volume} {32}},\ \bibinfo {pages} {446} (\bibinfo {year}
  {1936})}\BibitemShut {NoStop}%
\bibitem [{\citenamefont {Quintino}\ \emph {et~al.}(2015)\citenamefont
  {Quintino}, \citenamefont {V{\'{e}}rtesi}, \citenamefont {Cavalcanti},
  \citenamefont {Augusiak}, \citenamefont {Demianowicz}, \citenamefont
  {Ac{\'{\i}}n},\ and\ \citenamefont {N.}}]{Quintino_2015}%
  \BibitemOpen
  \bibfield  {author} {\bibinfo {author} {\bibfnamefont {M.~T.}\ \bibnamefont
  {Quintino}}, \bibinfo {author} {\bibfnamefont {T.}~\bibnamefont
  {V{\'{e}}rtesi}}, \bibinfo {author} {\bibfnamefont {D.}~\bibnamefont
  {Cavalcanti}}, \bibinfo {author} {\bibfnamefont {R.}~\bibnamefont
  {Augusiak}}, \bibinfo {author} {\bibfnamefont {M.}~\bibnamefont
  {Demianowicz}}, \bibinfo {author} {\bibfnamefont {A.}~\bibnamefont
  {Ac{\'{\i}}n}}, \ and\ \bibinfo {author} {\bibfnamefont {B.}~\bibnamefont
  {N.}},\ }\href {\doibase 10.1103/physreva.92.032107} {\bibfield  {journal}
  {\bibinfo  {journal} {Phys. Rev. A}\ }\textbf {\bibinfo {volume} {92}}
  (\bibinfo {year} {2015}),\ 10.1103/physreva.92.032107}\BibitemShut {NoStop}%
\bibitem [{Note1()}]{Note1}%
  \BibitemOpen
  \bibinfo {note} {The insphere of a polytope is the largest centered sphere
  contained in it.}\BibitemShut {Stop}%

 
 \bibitem{Bertlmann} R. A. Bertlmann and P. Krammer,Bloch vectors for qudits, Journal of Physics A: Mathematical and Theoretical \textbf{23}, 235303 (2008).
 
 \bibitem{Brand2005} F. Brand\~ao, Quantifying entanglement with witness operators, Phys. Rev. A \textbf{72}, 022310 (2005).

 \bibitem{Doherty1}A. C. Doherty, P. A. Parrilo, F. M Spedalieri, Distinguishing Separable and Entangled States, Phys. Rev. Lett. \textbf{88}, 187904 (2002)
 
  \bibitem{Doherty2}A. C. Doherty, P. A. Parrilo, F. M Spedalieri, Distinguishing Separable and Entangled States, Phys. Rev. A \textbf{69}, 022308 (2004)

\bibitem{guhne}B. Jungnitsch, T. Moroder, O. G\"uhne, Taming Multiparticle Entanglement, Phys. Rev. Lett. {\bf{106}}, 190502 (2011).

\bibitem [{Note2()}]{Note2}%
  \BibitemOpen
  \bibinfo {note} {In fact, \cite {Hirsch_2013} present a more general
  construction which works for LHV models. We only need the weaker result
  stated, which is implicit in their construction.}\BibitemShut {Stop}%
\bibitem [{\citenamefont {Hirsch}\ \emph {et~al.}(2015)\citenamefont {Hirsch},
  \citenamefont {Quintino}, \citenamefont {V\'ertesi}, \citenamefont {Pusey},\
  and\ \citenamefont {Brunner}}]{Geneva}%
  \BibitemOpen
  \bibfield  {author} {\bibinfo {author} {\bibfnamefont {F.}~\bibnamefont
  {Hirsch}}, \bibinfo {author} {\bibfnamefont {M.~T.}\ \bibnamefont
  {Quintino}}, \bibinfo {author} {\bibfnamefont {T.}~\bibnamefont {V\'ertesi}},
  \bibinfo {author} {\bibfnamefont {M.~F.}\ \bibnamefont {Pusey}}, \ and\
  \bibinfo {author} {\bibfnamefont {N.}~\bibnamefont {Brunner}},\ }\href@noop
  {} {\bibfield  {journal} {\bibinfo  {journal} {arXiv:1512.00262}\ } (\bibinfo
  {year} {2015})}\BibitemShut {NoStop}%
\bibitem [{\citenamefont {Christof}\ and\ \citenamefont
  {Loebel}(1997)}]{porta}%
  \BibitemOpen
  \bibfield  {author} {\bibinfo {author} {\bibfnamefont {T.}~\bibnamefont
  {Christof}}\ and\ \bibinfo {author} {\bibfnamefont {A.}~\bibnamefont
  {Loebel}},\ }\href@noop {} {\enquote {\bibinfo {title} {Porta: Polyhedron
  representation transformation algorithm},}\ }\bibinfo {howpublished}
  {\url{http://porta.zib.de}} (\bibinfo {year} {1997})\BibitemShut {NoStop}%
\bibitem [{\citenamefont {Avis}(2015)}]{lrs}%
  \BibitemOpen
  \bibfield  {author} {\bibinfo {author} {\bibfnamefont {D.}~\bibnamefont
  {Avis}},\ }\href@noop {} {\enquote {\bibinfo {title} {lrs, version 6.1},}\
  }\bibinfo {howpublished} {\url{http://cgm.cs.mcgill.ca/~avis/C/lrs.html}}
  (\bibinfo {year} {2015})\BibitemShut {NoStop}%
\bibitem [{\citenamefont {Pusey}(2013)}]{Pusey_2013}%
  \BibitemOpen
  \bibfield  {author} {\bibinfo {author} {\bibfnamefont {M.~F.}\ \bibnamefont
  {Pusey}},\ }\href {\doibase 10.1103/physreva.88.032313} {\bibfield  {journal}
  {\bibinfo  {journal} {Phys. Rev. A}\ }\textbf {\bibinfo {volume} {88}}
  (\bibinfo {year} {2013}),\ 10.1103/physreva.88.032313}\BibitemShut {NoStop}%
\bibitem [{\citenamefont {Nielsen}\ and\ \citenamefont
  {Chuang}(2011)}]{Nielsen:2011:QCQ:1972505}%
  \BibitemOpen
  \bibfield  {author} {\bibinfo {author} {\bibfnamefont {M.~A.}\ \bibnamefont
  {Nielsen}}\ and\ \bibinfo {author} {\bibfnamefont {I.~L.}\ \bibnamefont
  {Chuang}},\ }\href@noop {} {\emph {\bibinfo {title} {Quantum Computation and
  Quantum Information: 10th Anniversary Edition}}},\ \bibinfo {edition} {10th}\
  ed.\ (\bibinfo  {publisher} {Cambridge University Press},\ \bibinfo {address}
  {New York, NY, USA},\ \bibinfo {year} {2011})\BibitemShut {NoStop}%
\bibitem [{\citenamefont {Hofmann}\ \emph {et~al.}(2012)\citenamefont
  {Hofmann}, \citenamefont {Krug}, \citenamefont {Ortegel}, \citenamefont
  {G{\'e}rard}, \citenamefont {Weber}, \citenamefont {Rosenfeld},\ and\
  \citenamefont {Weinfurter}}]{Hofmann72}%
  \BibitemOpen
  \bibfield  {author} {\bibinfo {author} {\bibfnamefont {J.}~\bibnamefont
  {Hofmann}}, \bibinfo {author} {\bibfnamefont {M.}~\bibnamefont {Krug}},
  \bibinfo {author} {\bibfnamefont {N.}~\bibnamefont {Ortegel}}, \bibinfo
  {author} {\bibfnamefont {L.}~\bibnamefont {G{\'e}rard}}, \bibinfo {author}
  {\bibfnamefont {M.}~\bibnamefont {Weber}}, \bibinfo {author} {\bibfnamefont
  {W.}~\bibnamefont {Rosenfeld}}, \ and\ \bibinfo {author} {\bibfnamefont
  {H.}~\bibnamefont {Weinfurter}},\ }\href {\doibase 10.1126/science.1221856}
  {\bibfield  {journal} {\bibinfo  {journal} {Science}\ }\textbf {\bibinfo
  {volume} {337}},\ \bibinfo {pages} {72} (\bibinfo {year} {2012})},\ \Eprint
  {http://arxiv.org/abs/http://science.sciencemag.org/content/337/6090/72.full.pdf}
  {http://science.sciencemag.org/content/337/6090/72.full.pdf} \BibitemShut
  {NoStop}%
\bibitem [{\citenamefont {{Pironio}}\ \emph {et~al.}(2010)\citenamefont
  {{Pironio}}, \citenamefont {{Ac{\'{\i}}n}}, \citenamefont {{Massar}},
  \citenamefont {{de La Giroday}}, \citenamefont {{Matsukevich}}, \citenamefont
  {{Maunz}}, \citenamefont {{Olmschenk}}, \citenamefont {{Hayes}},
  \citenamefont {{Luo}}, \citenamefont {{Manning}},\ and\ \citenamefont
  {{Monroe}}}]{2010Natur.464.1021P}%
  \BibitemOpen
  \bibfield  {author} {\bibinfo {author} {\bibfnamefont {S.}~\bibnamefont
  {{Pironio}}}, \bibinfo {author} {\bibfnamefont {A.}~\bibnamefont
  {{Ac{\'{\i}}n}}}, \bibinfo {author} {\bibfnamefont {S.}~\bibnamefont
  {{Massar}}}, \bibinfo {author} {\bibfnamefont {A.~B.}\ \bibnamefont {{de La
  Giroday}}}, \bibinfo {author} {\bibfnamefont {D.~N.}\ \bibnamefont
  {{Matsukevich}}}, \bibinfo {author} {\bibfnamefont {P.}~\bibnamefont
  {{Maunz}}}, \bibinfo {author} {\bibfnamefont {S.}~\bibnamefont
  {{Olmschenk}}}, \bibinfo {author} {\bibfnamefont {D.}~\bibnamefont
  {{Hayes}}}, \bibinfo {author} {\bibfnamefont {L.}~\bibnamefont {{Luo}}},
  \bibinfo {author} {\bibfnamefont {T.~A.}\ \bibnamefont {{Manning}}}, \ and\
  \bibinfo {author} {\bibfnamefont {C.}~\bibnamefont {{Monroe}}},\ }\href
  {\doibase 10.1038/nature09008} {\bibfield  {journal} {\bibinfo  {journal}
  {\nat}\ }\textbf {\bibinfo {volume} {464}},\ \bibinfo {pages} {1021}
  (\bibinfo {year} {2010})},\ \Eprint {http://arxiv.org/abs/0911.3427}
  {arXiv:0911.3427 [quant-ph]} \BibitemShut {NoStop}%
\bibitem [{\citenamefont {Brand{\~{a}}o}(2005)}]{Brand_o_2005}%
  \BibitemOpen
  \bibfield  {author} {\bibinfo {author} {\bibfnamefont {F.~G. S.~L.}\
  \bibnamefont {Brand{\~{a}}o}},\ }\href {\doibase 10.1103/physreva.72.022310}
  {\bibfield  {journal} {\bibinfo  {journal} {Phys. Rev. A}\ }\textbf {\bibinfo
  {volume} {72}} (\bibinfo {year} {2005}),\
  10.1103/physreva.72.022310}\BibitemShut {NoStop}%
\bibitem [{\citenamefont {Eisert}\ \emph {et~al.}(2007)\citenamefont {Eisert},
  \citenamefont {Brand{\~{a}}o},\ and\ \citenamefont
  {Audenaert}}]{Eisert_2007}%
  \BibitemOpen
  \bibfield  {author} {\bibinfo {author} {\bibfnamefont {J.}~\bibnamefont
  {Eisert}}, \bibinfo {author} {\bibfnamefont {F.~G. S.~L.}\ \bibnamefont
  {Brand{\~{a}}o}}, \ and\ \bibinfo {author} {\bibfnamefont {K.~M.~R.}\
  \bibnamefont {Audenaert}},\ }\href {\doibase 10.1088/1367-2630/9/3/046}
  {\bibfield  {journal} {\bibinfo  {journal} {New J. Phys.}\ }\textbf {\bibinfo
  {volume} {9}},\ \bibinfo {pages} {46} (\bibinfo {year} {2007})}\BibitemShut
  {NoStop}%
\bibitem [{Note3()}]{Note3}%
  \BibitemOpen
  \bibinfo {note} {Typically this is convergence up to numerical precision.
  Note also that this can occur at a local maximum.}\BibitemShut {Stop}%
\bibitem [{Note4()}]{Note4}%
  \BibitemOpen
  \bibinfo {note} {The data will be maintained at \protect \url
  {https://git.io/vV7Bu}}\BibitemShut {NoStop}%
\bibitem [{\citenamefont {Jungnitsch}\ \emph {et~al.}(2011)\citenamefont
  {Jungnitsch}, \citenamefont {Moroder},\ and\ \citenamefont
  {G\"uhne}}]{Guhne}%
  \BibitemOpen
  \bibfield  {author} {\bibinfo {author} {\bibfnamefont {B.}~\bibnamefont
  {Jungnitsch}}, \bibinfo {author} {\bibfnamefont {T.}~\bibnamefont {Moroder}},
  \ and\ \bibinfo {author} {\bibfnamefont {O.}~\bibnamefont {G\"uhne}},\ }\href
  {\doibase 10.1103/PhysRevLett.106.190502} {\bibfield  {journal} {\bibinfo
  {journal} {Phys. Rev. Lett.}\ }\textbf {\bibinfo {volume} {106}},\ \bibinfo
  {pages} {190502} (\bibinfo {year} {2011})}\BibitemShut {NoStop}%
\bibitem [{\citenamefont {T\'oth}\ and\ \citenamefont
  {Ac\'{\i}n}(2006)}]{GezaToni}%
  \BibitemOpen
  \bibfield  {author} {\bibinfo {author} {\bibfnamefont {G.}~\bibnamefont
  {T\'oth}}\ and\ \bibinfo {author} {\bibfnamefont {A.}~\bibnamefont
  {Ac\'{\i}n}},\ }\href {\doibase 10.1103/PhysRevA.74.030306} {\bibfield
  {journal} {\bibinfo  {journal} {Phys. Rev. A}\ }\textbf {\bibinfo {volume}
  {74}},\ \bibinfo {pages} {030306} (\bibinfo {year} {2006})}\BibitemShut
  {NoStop}%
\bibitem [{\citenamefont {Bowles}\ \emph
  {et~al.}(2016{\natexlab{b}})\citenamefont {Bowles}, \citenamefont
  {Francfort}, \citenamefont {Fillettaz}, \citenamefont {Hirsch},\ and\
  \citenamefont {Brunner}}]{MultiGeneva}%
  \BibitemOpen
  \bibfield  {author} {\bibinfo {author} {\bibfnamefont {J.}~\bibnamefont
  {Bowles}}, \bibinfo {author} {\bibfnamefont {J.}~\bibnamefont {Francfort}},
  \bibinfo {author} {\bibfnamefont {M.}~\bibnamefont {Fillettaz}}, \bibinfo
  {author} {\bibfnamefont {F.}~\bibnamefont {Hirsch}}, \ and\ \bibinfo {author}
  {\bibfnamefont {N.}~\bibnamefont {Brunner}},\ }\href {\doibase
  10.1103/PhysRevLett.116.130401} {\bibfield  {journal} {\bibinfo  {journal}
  {Phys. Rev. Lett.}\ }\textbf {\bibinfo {volume} {116}},\ \bibinfo {pages}
  {130401} (\bibinfo {year} {2016}{\natexlab{b}})}\BibitemShut {NoStop}%
\bibitem [{\citenamefont {Slater}(2007)}]{Slater}%
  \BibitemOpen
  \bibfield  {author} {\bibinfo {author} {\bibfnamefont {P.~B.}\ \bibnamefont
  {Slater}},\ }\href {\doibase 10.1103/PhysRevA.75.032326} {\bibfield
  {journal} {\bibinfo  {journal} {Phys. Rev. A}\ }\textbf {\bibinfo {volume}
  {75}},\ \bibinfo {pages} {032326} (\bibinfo {year} {2007})}\BibitemShut
  {NoStop}%
\bibitem [{Note5()}]{Note5}%
  \BibitemOpen
  \bibinfo {note} {All the SDPs implemented in this work used {\protect \sc
  matlab} and the packages {\protect \sc cvx} \cite {cvx,gb08} and {\protect
  \sc qetlab} \cite {qetlab}.}\BibitemShut {Stop}%
\bibitem [{\citenamefont {Grant}\ and\ \citenamefont {Boyd}(2014)}]{cvx}%
  \BibitemOpen
  \bibfield  {author} {\bibinfo {author} {\bibfnamefont {M.}~\bibnamefont
  {Grant}}\ and\ \bibinfo {author} {\bibfnamefont {S.}~\bibnamefont {Boyd}},\
  }\href {http://cvxr.com/cvx} {\enquote {\bibinfo {title} {{{CVX}: Matlab
  Software for Disciplined Convex Programming, version 2.1}},}\ }\bibinfo
  {howpublished} {\url{http://cvxr.com/cvx}} (\bibinfo {year}
  {2014})\BibitemShut {NoStop}%
\bibitem [{\citenamefont {Grant}\ and\ \citenamefont {Boyd}(2008)}]{gb08}%
  \BibitemOpen
  \bibfield  {author} {\bibinfo {author} {\bibfnamefont {M.}~\bibnamefont
  {Grant}}\ and\ \bibinfo {author} {\bibfnamefont {S.}~\bibnamefont {Boyd}},\
  }in\ \href {http://stanford.edu/~boyd/graph_dcp.html} {\emph {\bibinfo
  {booktitle} {Recent Advances in Learning and Control}}},\ \bibinfo {series
  and number} {Lecture Notes in Control and Information Sciences},\ \bibinfo
  {editor} {edited by\ \bibinfo {editor} {\bibfnamefont {V.}~\bibnamefont
  {Blondel}}, \bibinfo {editor} {\bibfnamefont {S.}~\bibnamefont {Boyd}}, \
  and\ \bibinfo {editor} {\bibfnamefont {H.}~\bibnamefont {Kimura}}}\ (\bibinfo
   {publisher} {Springer-Verlag Limited},\ \bibinfo {year} {2008})\ pp.\
  \bibinfo {pages} {95--110}\BibitemShut {NoStop}%
\bibitem [{\citenamefont {Johnston}(2015)}]{qetlab}%
  \BibitemOpen
  \bibfield  {author} {\bibinfo {author} {\bibfnamefont {N.}~\bibnamefont
  {Johnston}},\ }\href {\doibase 10.5281/zenodo.14186} {\enquote {\bibinfo
  {title} {{{QETLAB}: A {MATLAB} toolbox for quantum entanglement, version
  0.8}},}\ }\bibinfo {howpublished} {\url{http://qetlab.com}} (\bibinfo {year}
  {2015})\BibitemShut {NoStop}%
\end{thebibliography}
%

\end{document}